\documentclass[preprint,a4paper,11pt,authoryear]{elsarticle}
\usepackage[margin=1in]{geometry}
\usepackage{natbib}
\usepackage{amsmath}
\usepackage{adjustbox}
\usepackage{graphicx}
\usepackage{multirow}
\usepackage{amsmath, amssymb}
\usepackage{hhline}
\usepackage{array}
\usepackage{amsfonts}
\usepackage{xcolor}
\usepackage{hyperref}
\usepackage{cancel}
\usepackage{enumitem}
\newtheorem{theorem}{Theorem}
\newtheorem{proposition}[theorem]{Proposition}
\newcommand{\killpunct}[1]{}
\usepackage{enumitem}

\title{Computational adversarial risk analysis for general security games}

\author[inst1]{J.M. Camacho \corref{cor1}}
\author[inst2]{R. Naveiro}
\author[inst1]{D. Rios Insua}

\address[inst1]{ICMAT-CSIC, Madrid, Spain}
\address[inst2]{CUNEF Universidad, Madrid, Spain}

\cortext[cor1]{Corresponding author: josemanuel.camacho@icmat.es}

\newcommand{\Acal}{\mathcal{A}}
\newcommand{\Dcal}{\mathcal{D}}

\newcommand{\Pcal}{\mathcal{P}}

\newcommand{\Pbb}{\mathbb{P}}
\newcommand{\opt}{*}
\newcommand{\dd}{\mathop{}\! \mathrm{d}}
\DeclareMathOperator*{\mode}{mode}
\usepackage[official]{eurosym}
\usepackage{subcaption} 
\usepackage[boxed, vlined]{algorithm2e}
\SetKwComment{Comment}{$\triangleright$\ }{} %
\SetAlCapSkip{0.5em}
\SetKwInput{Input}{input}
\SetKwInOut{Initialize}{initialize}
\SetEndCharOfAlgoLine{}
\DeclareMathOperator*{\argmax}{arg\,max}

\SetKwIF{If}{ElseIf}{Else}{if}{}{else if}{else}{end if}%
\SetKwFor{While}{while}{}{end while}%
\SetKwRepeat{Do}{do}{while}
\SetKw{KwGoTo}{go to}

\usepackage{setspace}
\makeatletter
\def\ps@pprintTitle{%
  \let\@oddhead\@empty
  \let\@evenhead\@empty
  \let\@oddfoot\@empty
  \let\@evenfoot\@empty
}
\makeatother

\begin{document}

\begin{frontmatter}

\begin{abstract}
 This paper provides an efficient computational scheme to handle general security games from an adversarial risk analysis perspective. Two cases in relation to single-stage and multi-stage simultaneous defend-attack games motivate our approach to general setups %
   which uses bi-agent influence diagrams as underlying problem structure and augmented probability simulation as core computational methodology. Theoretical convergence and 
 numerical, modeling, and implementation issues are thoroughly discussed. A disinformation war case study illustrates the relevance of the proposed approach.
\end{abstract}
\end{frontmatter}

\section{Introduction}
\label{sec:intro}

\noindent Security games provide a powerful and flexible modeling framework
for strategic and operational defense and homeland security
  (DHS)
problems, as  
\citet{Brown2006}, \citet{Zhuang2007}, %
 or \citet{hausken2024fifty} cogently argue.
These authors illustrate their analysis mainly
from a standard game-theoretic perspective which, 
 computationally, is based on approximating Nash
equilibria and related refinements.
Within them, it is
frequently assumed that agents know not only their own
payoffs, preferences, beliefs, and feasible actions, but also those of their opponents.
Alternatively, in games with incomplete information \citep{Harsanyi1967}, it is typically considered that each agent has a probability distribution over their opponents' types which  characterize their beliefs and preferences. This distribution is assumed to be known by all the players, the {\em common prior hypothesis}, which allows for a symmetric joint normative analysis of the game in which players maximise expected utilities, and expect other players to do the same. %
General {\em common knowledge } assumptions underlying these
approaches have been critically reviewed in e.g. \citet{Raiffa2002} or \citet{HargreavesHeap1995}, whereas common prior assumptions are 
discussed by \cite{antos2010representing}, \cite{sakovics2001games} or \cite{angeletos2018forward}, to name but a few.
In the DHS domain, 
agents will generally lack such common knowledge as 
adversaries try to hide information
and even disinform opponents.

Adversarial Risk Analysis (ARA)  \citep{RiosInsua2009} provides an alternative solution framework partly mitigating such common knowledge assumptions. %
ARA supports one of the decision-makers (designated 
defender) seeking for actions that maximize her expected utility. However, procedures employing the game theoretic structure are 
incorporated to predict %
the opponents' (designated attackers) actions.  
ARA thus makes operational the Bayesian approach to games, as 
sketched in \citet{Kadane1982}
or \citet{Raiffa2002}. \cite{banks2022adversarial} provide an in-depth comparison of ARA with standard game-theoretic approaches. In particular, different attacker rationalities may be considered in such framework. %
  Our focus will be on the scenario of a defender facing a single attacker. Additionally, we constrain the defender to act as a level-2 thinker in the sense of \citet{Stahl1995}: the attacker is assumed to maximize expected utility while modeling the defender as a non-strategic player. However, the defender's uncertainty regarding the attacker's judgments propagates to his optimal decision, that becomes random and results in a probabilistic prediction of the attack.

A main motivation for ARA comes from the DHS realm. 
Since its proposal, it has been used to model a variety of problems including  
anti-IED war \citep{wang11}, international  piracy \citep{sevillano12}, %
counter-terrorism  online surveillance \citep{Gil19}, cyber-security \citep{rios19}, and 
combat modeling enhancement \citep{roponen2020adversarial} scenarios.
Note though that its computational requirements 
  are non-negligible since it essentially entails alternating 
    stages in which we simulate from the attacker's problem 
to produce a probabilistic forecast of the adversary's actions and then optimize within the defender's problem, with such forecast as input. 

From a methodological point of view, in his celebrated paper, \cite{Shachter1986} proposed as a major open problem devising methods to solve influence diagrams (ID) with multiple decision-makers. \cite{Koller2003} provided an algorithm to compute Nash equilibria in discrete problems under common knowledge assumptions; \cite{virtanen2006modeling} extended the ideas to sequential parallel games computing Nash equilibria in air combat models;  \cite{BAIDS} provided an ARA solution in discrete problems. In turn, \cite{ekin2023augmented} explored  
augmented probability simulation 
(APS, \citealt{BMI:1999}) as a solution method to handle simple sequential single-stage defend-attack games from an ARA perspective.
We show here how APS may be used to solve general
security games described through, possibly 
continuous, bi-agent influence 
diagrams (BAID) from an ARA perspective, therefore
providing the first general solution to Shachter's query. 

For this, we initially describe how to deal with simultaneous defend-attack (Section \ref{sec:sec_template}) games and their sequential versions.
  Such streamlined security models serve to motivate an efficient 
computational method to handle general security games detailed in Section \ref{sec:m_s_gsg}. 
Section \ref{sec:computational_issues} presents key
numerical strategies to support the implementation of our methods. A disinformation war example then illustrates the developments in Section \ref{sec:casestudy}.
We end up with a brief discussion. Proofs of key propositions are given in  \ref{sec:app_proof_pro}.  Software 
 to reproduce the example is available at \href{https://github.com/jmcamachor1/general\_ARA\_APS}{https://github.com/jmcamachor1/general\_ARA\_APS}. 
 Supplementary materials provide details of the parametric choices in the case and how we solved it, and proofs of the remaining propositions.

\section{Solving simultaneous Defend-Attack games} 
\label{sec:sec_template}
This section introduces key notation and sketches 
how APS-based methods find optimal defenses in the important class of multiple-stage simultaneous defend-attack games under incomplete information. It draws on \citet{ekin2023augmented} who handled two-stage sequential defend-attack games.

\subsection{Single-stage simultaneous defend-attack games}
\label{sec:single_stage_parallel}

\begin{figure}[htbp]
\begin{subfigure}[t]{0.32\textwidth}
\centering
\includegraphics[width=0.65\linewidth]{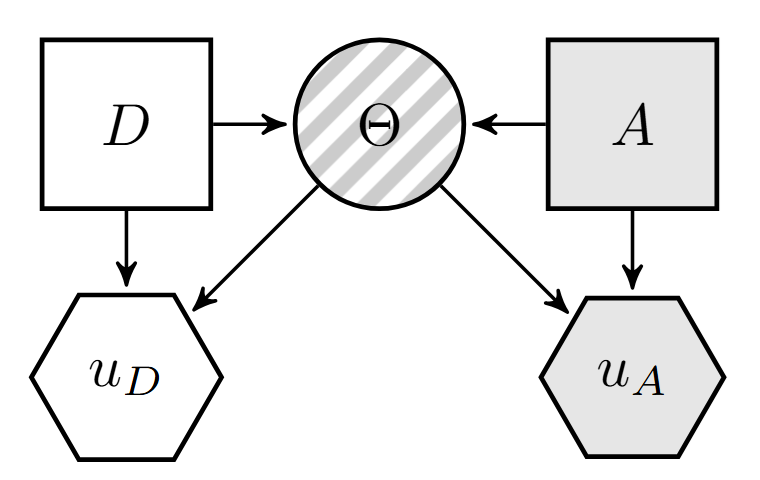} 
\caption{}
\end{subfigure}
\hfill
\begin{subfigure}[t]{0.32\textwidth}
\centering
\includegraphics[width=0.65\linewidth]{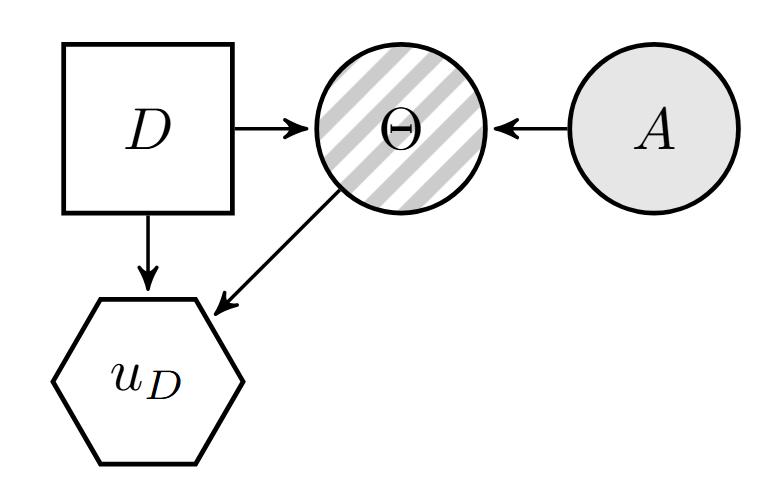} 
\caption{}
\end{subfigure}
\hfill
\begin{subfigure}[t]{0.32\textwidth}
\centering
\includegraphics[width=0.65\linewidth]{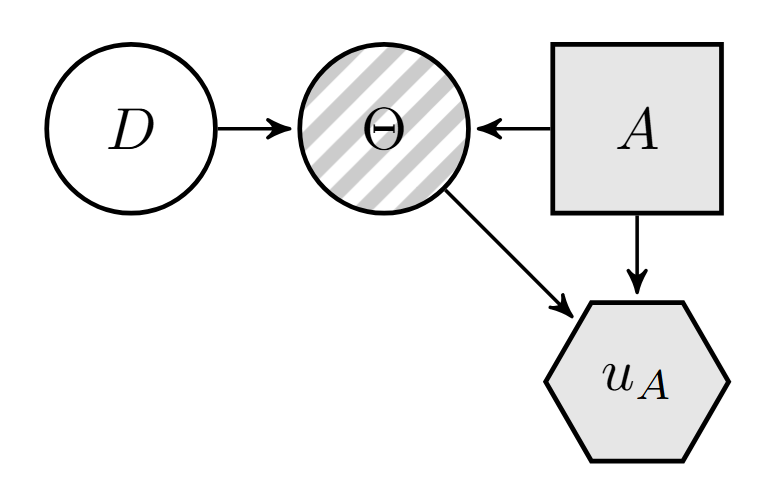} 
\caption{}
\end{subfigure}
\caption{ (a) Simultaneous defend-attack game BAID, (b) D's decision problem, (c) D's analysis of A's problem.}
\label{fig:gdp}
\end{figure}

Assume a Defender ($D$, she) who has to choose her defense $d$ from $\Dcal$, her set of feasible alternatives. In parallel, an Attacker ($A$, he) chooses his attack $a$ from his feasible set $\Acal$. The consequences of the interaction for both agents depend on a random outcome $\theta \in \Theta$.
The agents have their own assessment of the random outcome's probability, respectively modeled through $p_D(\theta \vert d,a)$ and $p_A(\theta  \vert d,a)$. $D$'s utility function $u_D(d, \theta)$ depends on her chosen defense and the outcome; similarly, $A$'s utility function has the form $u_A(a, \theta)$. Such simultaneous defend-attack game is depicted as a BAID \citep{banks16} in Figure \ref{fig:gdp}a. The diagram shows two decision nodes (squares), one chance node (circle), and two utility nodes (hexagons), jointly representing the problems faced by the Defender (white) and the Attacker (grey). Striped nodes refer 
 to chance events pertinent to both agents' decisions. Upon examining the Defender's problem (Figure \ref{fig:gdp}b), the Attacker's decision $A$ appears as an uncertainty 
 from $D$'s viewpoint. To address her problem, the Defender analyzes the situation from \emph{A}'s standpoint (Figure \ref{fig:gdp}c). In this case, node $D$ in Figure \ref{fig:gdp}a becomes a chance node for the Attacker, showing his uncertainty regarding the Defender's intentions. %

\noindent

\noindent As discussed, the complete information assumption will not hold in many security scenarios which we shall handle with a decision-analytic approach. %
  Let $D$'s expected utility  be
$\psi_D(d) = \int \psi_D(d,a)\, p_D(a) \dd a $, where $\psi_D (d,a) = \int u_D (d, \theta )\, p_D(\theta \vert d,a) \dd \theta $.\footnote{We assume here that $A$, $D$ and $\Theta$ are all continuous. Discrete cases are handled in a similar manner.}  Her optimal decision will be $d^\opt_\text{ARA} = \argmax_{d \in \Dcal}\, \psi_D(d)$. Finding 
 it requires estimating $p_D(a)$, $D$'s assessment of $A$ choosing attack $a$. For this, consider $A$'s problem who aims to find
 $a^* = \argmax_{a \in \Acal} \int \psi_A(d, a)  p_A(d) \dd d,
$ with $\psi_A(d,a)=\int u_A(a,\theta)p_A(\theta|d,a)\dd \theta.$
However, as \cite{KEENEY} argues, $D$ has no access to 
   $A$ to assess $u_A$ and $p_A$, and, therefore, 
   lacks complete knowledge about $A$'s utilities and probabilities.
   Adopting a Bayesian approach, let us  
  model such uncertainty through random utilities  $U_A(a, \theta)$ and random probabilities $P_A(\theta | d, a)$ and $P_A (d)$, defined over a common probability space $(\Omega, \cal F, \Pcal)$ with atomic elements $\omega\in \Omega$ \citep{Chung}. Then, 
the random optimal attack is
$ 
A^*  = \argmax_{x \in \Acal} \int \Psi_A (d,x) \cdot P_A  (d) \dd d, 
$ 
with $\Psi_A(d,x) = \int U_A(x,\theta)P_A(\theta|d,x)\dd \theta$, and we make $p_D(A\leq a) =  Pr \left[ A^* \leq a  \right]= {\cal P} \{ \omega :  A^{*\omega } 
\leq  a  \}$. %
This defines $p_D(a)$, the
missing ingredient when searching for $d^*_{ARA}$. Implementing this requires the specification of $U_A(a, \theta)$, $P_A(\theta | d,a)$, and $P_A(d)$. Of these, eliciting $P_A(d)$ has a recursive element as we need to think about how the attacker reasons about the defender which, in turn, begs for thinking about 'how the defender thinks about how the attacker thinks about the defender', and so on. This leads to a recursion similar to the level-$k$ scheme described in \citet{rios2012adversarial}. 
As mentioned, we only consider level-2 defenders,
although the scheme described extends to higher levels in the thinking hierarchy.

 As seen,
ARA models entail integration and optimization procedures that may be computationally challenging. This paper proposes efficient APS-based methods to handle them. First, for the Defender problem, assuming with no loss of generality that $u_D$ is positive, we introduce an augmented distribution (AD) over $(d,a, \theta)$ for $D$'s problem defined through 
$
\pi_D(d, a, \theta) \propto u_D(d,\theta) \cdot p_D(\theta \vert d,a)\cdot p_D(a)$.
Observe that its marginal $\pi_D(d)$ in $d$ is $\iint u_D (d, \theta) \cdot p_D(\theta \vert d, a)\cdot p_D(a) \dd \theta \dd a$ which is proportional to  $\psi_D(d)$. Consequently, $d^\opt_\text{ARA}  = \mode{(\pi_D(d))}$. In such a way, one would just need to sample $(d, a, \theta )\sim \pi_D(d, a, \theta)$ and estimate its mode in $d$ to approximate $d^*_{ARA}$.  Sampling may be performed using Markov Chain Monte Carlo (MCMC) methods. In particular, Algorithm \ref{alg:ARA_APS_framework}b) ($\texttt{DAPS}$ function) illustrates a Metropolis-Hastings (MH) \citep{French:2000} version.  Besides the Defender's utility $u_D$ and probabilities $p_D$, \texttt{DAPS} takes as input the number $N$ of MCMC samples and the candidate generating distribution $g_D$ within the function $\texttt{MH}$. 

{\small
\begin{algorithm}[H]
\textcolor{black}{
\linespread{0.7}\selectfont
\SetKwFunction{Fatk}{MH}
\SetKwProg{Fn}{function}{:}{}
\Fn{\Fatk{$g$, $u$, $x$, $x'$, $y$, $y'$ $\theta$, $\theta '$}}{
Compute acceptance probability $
    \alpha = \min \left \lbrace 1, \frac{u \left( x, \theta \right) \cdot g\left(x' \vert x \right) }{u \left( x', \theta ' \right) \cdot g\left( x \vert x' \right) } \right \rbrace
    $\;
    $\text{ Set  $z$= }\begin{cases}  (x,y, \theta)    & $\text{with probability $\alpha$ }$  \\  (x',y', \theta')    & $\text{with probability $1-\alpha$ }$  \end{cases}$\;
\textbf{return} $z$ \;}
\;
\linespread{0.7}\selectfont
\SetKwFunction{Fatk}{DAPS}
\SetKwProg{Fn}{function}{:}{}
\Fn{\Fatk{$N$, $u_D$, $p_D$, $g_D$}}{
\Initialize{ $d^{(0)}$  }
Draw $a^{(0)} \sim p_D( a )$\;
Draw $\theta^{(0)} \sim p_D(\theta \vert d^{(0)}, a^{(0)})$\;
\For{$i=1$ \KwTo $N$} {
    Propose new defense $\tilde{d} \sim g_D(\tilde{d} \vert d^{(i-1)})$\;
    Draw $\tilde{a} \sim p_D( a )$\; 
    Draw $\tilde{\theta} \sim p_D(\theta \vert \tilde{d}, \tilde{a} )$\;
    $(d^{(i)} , a^{(i)}, \theta^{(i)})$ = \texttt{MH}$(g_D,u_D,\tilde{d},d^{(i-1)}, \tilde{a},a^{(i-1)},\tilde{\theta},\theta^{(i-1)})$\;
    }
Discard burn-in samples and construct $\widehat{d}^\opt_\text{ARA}$, a consistent estimator of the mode of the marginal augmented distribution, using the remaining $d$ samples\;
\textbf{return} $\widehat{d}^\opt_\text{ARA}$ \;}
\;
\SetKwFunction{Fatk}{AAPS}
\SetKwProg{Fn}{function}{:}{}
\Fn{\Fatk{$M$, $U_A$, $P_A$, $g_A$}}{
\Initialize{ $a^{(0)}$}
Draw $u_A(a, \theta) \sim U_A(a,\theta)$\;
Draw $p_A(\theta \vert d,a) \sim P_A(\theta \vert d,a)$\;
Draw $p_A(d) \sim P_A(d)$\;
Draw $d^{(0)} \sim p_A(d)$\;
Draw $\theta^{(0)} \sim p_A(\theta \vert d^{(0)}, a^{(0)})$\;
\For{$i=1$ \KwTo $M$} {
	Propose new attack $\tilde{a} \sim g_A(\tilde{a} \vert a^{(i-1)})$ \;
	Draw $\tilde{d} \sim p_A(d)$\;
	Draw $\tilde{\theta} \sim p_A(\theta \vert \tilde{d}, \tilde{a})$\;
    $( a^{(i)}, d^{(i)}, \theta^{(i)})$ = \texttt{MH}$(g_D,u_D,\tilde{a},a^{(i-1)},\tilde{d},d^{(i-1)},\tilde{\theta},\theta^{(i-1)})$ \;
    }
Discard burn-in samples and construct $a^\opt$, a consistent estimator of the mode of the marginal augmented distribution, using the remaining $ a $ samples\;
\textbf{return} $a^\opt$\;
}}
\caption{\small a) \texttt{MH}: MH algorithm. b) \texttt{DAPS}: MH Defender APS to approximate\textcolor{white}{'}ARA solution. c) \texttt{AAPS}: MH Attacker APS to approximate sample from $p_D(a)$. }\label{alg:ARA_APS_framework}
\end{algorithm}
}

\vspace{0.25cm}

 $\texttt{DAPS}$ requires the ability to sample from $a \sim p_D(a)$. To produce such samples, assuming that $U_A^{\omega}$ is 
a.s.\ positive, consider the attacker random AD (RAD) model, defined for each $\omega \in \Omega$ through,
$ \Pi_A^{\omega}(d, \theta,a) \propto  U^{\omega}_{A}(a, \theta)\cdot P^{\omega}_A(\theta \vert d, a)\cdot P ^{\omega}_A(d)$.
Then, as before, a.s$.$ the mode of the marginal of $\Pi^\omega(d,\theta,a)$ in $a$ coincides with  $A^{*\omega}$. Consequently, by sampling $u_A(a, \theta) \sim U_A(a, \theta)$, $p_A(\theta \vert d, a) \sim P_A(\theta \vert d, a)$, and $p_A(d) \sim P_A(d)$, one can build $\pi_A(d, \theta, a) \propto u_A(a, \theta)p_A(\theta \vert d, a)p_A(d)$ which conforms a sample from $\Pi_A (d, \theta, a)$ (Algorithm \ref{alg:ARA_APS_framework}c). Therefore, $\text{mode}(\pi_A(a))$ is a sample from $A^*$,  thus providing a mechanism ($\texttt{AAPS}$) to sample from the target distribution. A repeated application of \texttt{AAPS} (Algorithm \ref{alg:ARA_APS_framework}c) would provide a sample of the required size from $A^{*}$. 
 Input to \texttt{AAPS} includes the number $M$ of samples, the candidate generating distribution $g_A$ for the algorithm, and the 
 Attacker's random utility $U_A$ and probabilities $P_A$.

Under mild conditions, the convergence of Algorithm \ref{alg:ARA_APS_framework}b 
to $d_{ARA }^*$ (supported by that of Algorithm \ref{alg:ARA_APS_framework}c) follows when using consistent mode estimators,
 as  proved in \ref{sec:app_proof_pro}.

\begin{proposition}
Suppose that
\begin{enumerate}
\item $u_D$ and, a.s., the utilities in the support of $U_A$ are positive and, respectively, continuous in $(d,\theta)$ and $(a, \theta)$.
\item The Defender's and Attacker's decision sets are compact.
 \item $p_D(\theta|d,a)$ and, a.s., the distributions in the support of $P_A(\theta|d,a)$ are continuous in $a$ and $d$, respectively, and positive in $(d,a)$.
 \item The proposal generating distributions $g_D$ and $g_A$ have, respectively, support $\mathcal{D}$ and $\mathcal{A}$.
\end{enumerate}

\noindent Then, Algorithm \ref{alg:ARA_APS_framework}b defines a Markov chain with stationary distribution $\pi_D(d, \theta, a)$. Moreover, a consistent mode estimator based on the marginal samples of $d$ from this Markov chain a.s.\ approximates $d^{*}_{ARA}$.
\end{proposition}

\noindent Observe that Algorithm \ref{alg:ARA_APS_framework} constructs two APS: one for the Defender problem (function $\texttt{DAPS}$), and another for the Attacker ($\texttt{AAPS}$), which is iteratively called.

\subsection{$n$-stage simultaneous defend-attack games}
\label{sec:n_stage_parallel}

\begin{figure}[htbp!]
  \centering
  \includegraphics[width=0.36\linewidth]{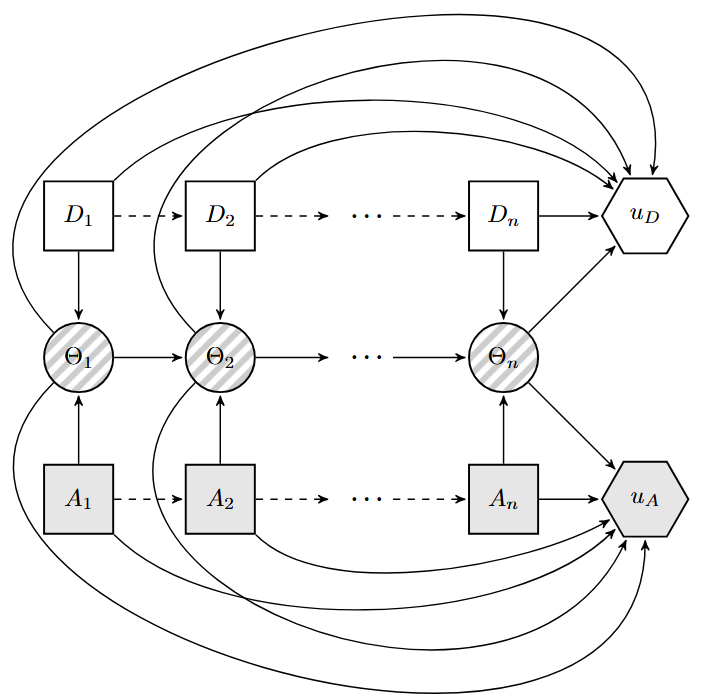} 
  \caption{{\small Template for basic \textit{n}-stage simultaneous defend-attack game BAID.}}
  \label{fig:n_sim_basic}
\end{figure}

Figure \ref{fig:n_sim_basic} depicts 
the case of $n$ stages with simultaneous decisions of both agents,
 with dashed arrows pointing to decision nodes indicating that 
 the corresponding decision is made
  knowing the values of its antecessors.
Numerous DHS scenarios, such as the air-combat model presented in \cite{virtanen2006modeling}, can be readily adapted to this framework. Likewise, reinforcement learning problems under threats, 
   introduced in \cite{gallego2019reinforcement}, can be reformulated to fit within this framework.

\begin{figure}[htbp]
\begin{subfigure}[t]{0.35\textwidth}
\centering
\raisebox{0.7cm}{\includegraphics[width=0.70\linewidth]{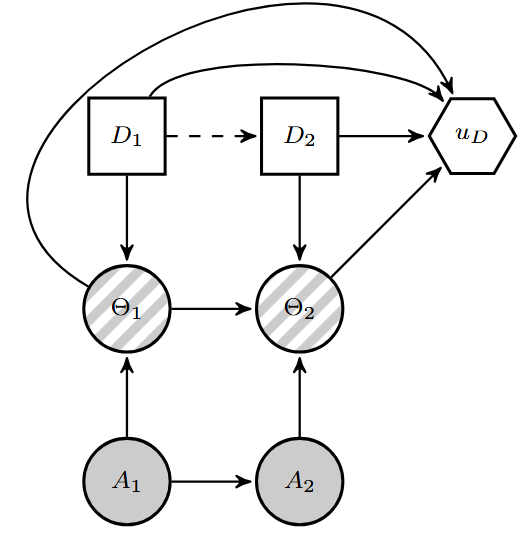}}
\caption{$D$'s influence diagram.}\label{AAA}
\end{subfigure}
\hfill
\begin{subfigure}[t]{0.49\textwidth}
\centering
  \includegraphics[width=0.50\linewidth]{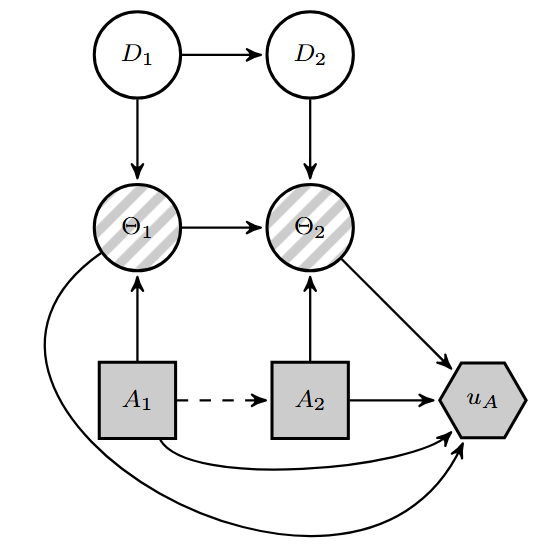} 
\caption{$A$'s influence diagram.}\label{BBB}
\end{subfigure}
\caption{\small{$D$'s and $A$'s influence diagrams for the 2-stage simultaneous defend-attack game.}}
\label{fig:n_sim_DA_ids}
\end{figure}

\pagebreak
Let us briefly discuss how the single-stage approach in Section \ref{sec:single_stage_parallel} extends to a 2-stage case. Figure \ref{fig:n_sim_DA_ids} presents the (a) defender's and (b) attacker's problems in this case. Assuming that the utility function $u_D(d_1,d_2,\theta_1,\theta_2)$, and the distributions $p_D(\theta_2|d_2,\theta_1, a_2)$, $p_D(a_2|a_1)$, $p_D(a_1)$, and $p_D(\theta_1|d_1,a_1)$ are available, we
 proceed as follows, using \cite{Shachter1986}'s algorithm to 
  solve the Defender's problem.

\begin{enumerate}[noitemsep]
    \item Remove chance node $\theta_2$, computing  $\psi_D(d_1,d_2,\theta_1, a_2)$=$\int$$u_D(d_1,d_2,\theta_1,\theta_2)p_D(\theta_2|d_2,\theta_1,a_2)$$\dd\theta_2$.
    \item Remove chance node $A_2$, computing 
    $\psi_D(d_1,d_2,\theta_1,a_1) = \int \psi_D(d_1,d_2,\theta_1,a_2)p_D(a_2|a_1)\dd a_2$.
    \item Remove chance node $\theta_1$, computing  $\psi_D(d_1,d_2,a_1) = \int \psi_D(d_1,d_2,\theta_1,a_1)p_D(\theta_1|a_1,d_1) \dd \theta_1$.
    \item Remove chance node $A_1$, computing $ \psi_D(d_1,d_2) = \int \psi_D(d_1,d_2,a_1)p_D(a_1)\dd a_1.$
    \item Remove decision node $D_2$, computing  $\psi_D(d_1) = \max_{d_2}\psi_D(d_1,d_2)$ and storing   $d_2^{*}(d_1) = \\\argmax_{d_2}\psi_D(d_1,d_2)$. 
    \item Remove decision node $D_1$, computing $d_1^{*} = \argmax_{d_1} \psi_D(d_1)$.
\end{enumerate}

\vspace{-0.25cm}

\paragraph{Remark 1} {\em We could combine steps 5 and 6, jointly removing nodes $D_1$ and $D_2$, to solve for $(d_1^{*},d_2^{*}) = \argmax_{(d_1,d_2)} \psi_D(d_1,d_2)$. However in some contexts it is of interest to provide a full second-stage policy $d_2^{*}(d_1)$. In particular, this will be relevant when addressing A's problem.} $ \hfill \triangle$
\vspace{.05in}

\pagebreak
\noindent  Interestingly, we may implement the above six steps through two DAPS reductions:
\begin{itemize}[noitemsep,label={},leftmargin=0pt]
\item {\em DAPS1}. Since  $ d_2^{*}(d_1)$=$\argmax_{d_2}\iiiint u_D(d_1,d_2,\theta_1,\theta_2)  p_D(\theta_2|d_2,\theta_1,a_2)p_D(a_2|a_1)p_D(\theta_1|d_1,a_1)\\ p_D(a_1)\dd a_1 \dd \theta_1 \dd a_2 \dd \theta_2$, we aggregate steps 1-5 by defining the AD 
 $\pi(d_2,\theta_1,\theta_2,a_1,a_2|d_1)\propto u_D(d_1,d_2,\theta_2)p_D(\theta_2|d_2,\theta_1,a_2)p_D(a_2|a_1)p_D(\theta_1|d_1,a_1)p_D(a_1)$,
        so that $\text{mode}(\pi(d_2|d_1)) = d_2^{*}(d_1)$. Next, proceed by simulating from $\pi(d_2,\theta_1,\theta_2,a_1,a_2|d_1)$ and, based on a sample from it, obtain the corresponding sample from $\pi(d_2|d_1)$, using it to get a consistent estimate of its mode.   

   \item {\em DAPS2.} This handles step 6 and uses $d_1^{*}=\argmax_{d_1}\psi_D(d_1)$. Although 
    obvious,\footnote{It is operationally obvious but computationally convenient to provide, later on, a unified approach, when facing general problems.} we view it through an AD with $\pi(d_1)\propto\psi_D(d_1)$ from which we sample to estimate its mode $\hat{d_1^{*}}$.
\end{itemize}

  As in Section \ref{sec:single_stage_parallel}, whereas $u_D(d_1,d_2,\theta_1, \theta_2)$, $p_D(\theta_2|d_2, \theta_1,a_2)$, and $p_D(\theta_1|d_1,a_1)$ are standard to elicit from a decision analytic point of view, $p_D(a_2|a_1)$ and $p_D(a_1)$ include strategic elements that need to be somehow reflected. For this, let us deal with the attacker problem from the defender point of view (Figure \ref{fig:n_sim_DA_ids}b), which is symmetric to her problem. %
    If, as in Section 2.1, we model our uncertainty about its ingredients with a random utility function $U_A$ and random probability distributions $P_A$, we would solve the attacker problem through two AAPS reductions:
  
\begin{itemize}[noitemsep,label={},leftmargin=0pt]
\item {\em AAPS1.} Define the RAD
$\Pi(a_2,\theta_1,\theta_2,d_1,d_2|a_1) \propto U_A(a_1,a_2,\theta_1,\theta_2) P_A(\theta_2|a_2,\theta_1,d_2)P_A(d_2|d_1)
\allowbreak P_A(\theta_1|a_1,d_1)P_A(d_1) $, 
to estimate the random optimal decision $\hat{A}_2^{*}(a_1)$. From it, estimate $\hat{p}_D(a_2|a_1)$.

\item {\em AAPS2.} Define the RAD $\Pi(a_1)\propto \Psi(a_1)$. From it sample the random optimal attack $\hat{A}_1^{*}$, and deduce $\hat{p}_D(a_1)$.\footnote{For completeness, we describe  estimations of $\hat{p}_D(a_2|a_1)$ and $\hat{p}_D(a_1)$ but,
 effectively, we only require samples from them.}
\end{itemize}

\noindent The approach is actually general as Proposition \ref{prop:n_sim} states. We omit its proof since it is a particular case of the more general Proposition \ref{prop:gen_algo} below.

\begin{proposition}
\label{prop:n_sim}
An n-stage simultaneous game can be solved from an ARA perspective by using n DAPS reductions and $n$ AAPS reductions.     
\end{proposition}

\section{General security games}
\label{sec:m_s_gsg}

The solution process sketched in Section \ref{sec:sec_template} 
  for specific types of security games will be adapted, formalized, and expanded to solve general ones, providing a 
 broad methodology to handle them based on ARA and APS. We 
   present the complexities of solving such security games through a disinformation war case,  later utilized to motivate and illustrate the methodology.

\paragraph{{\bf Case}} According to the Global Risks Map report \citep{world2024global,world2025global}, misinformation and disinformation will pose severe global risks in the short term, a situation further exacerbated by recent technological advances in e.g.\ large language models \citep{BARMAN2024100545}. As the reports highlight, this threat potentially entail multiple impacts
 affecting DHS, the environment, or public health, just to quote a few.
 
\begin{figure}[htbp]
\begin{subfigure}[t]{0.32\textwidth}
\centering
\includegraphics[width=0.90\linewidth]{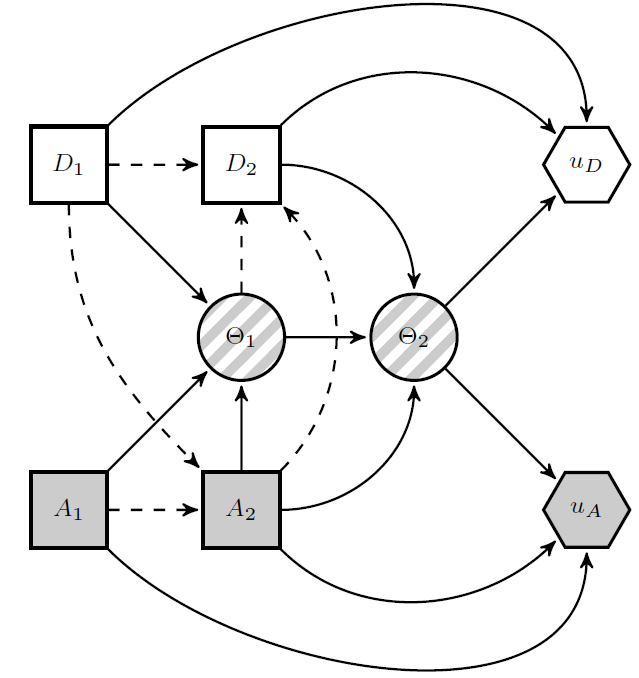}
\caption{}
\end{subfigure}
\hfill
\begin{subfigure}[t]{0.32\textwidth}
\centering
\raisebox{0.8cm}{
  \includegraphics[width=0.95\linewidth]{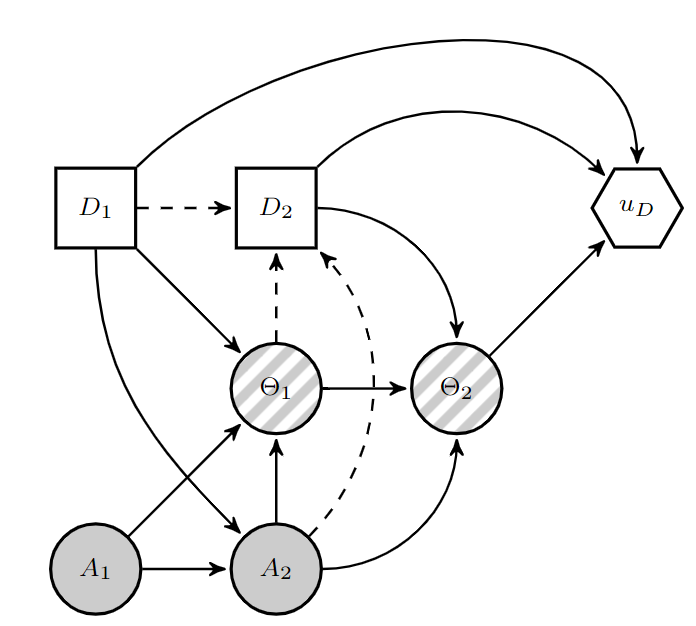}}
\caption{}
\end{subfigure}
\hfill
\begin{subfigure}[t]{0.32\textwidth}
\centering
  \includegraphics[width=0.95\linewidth]{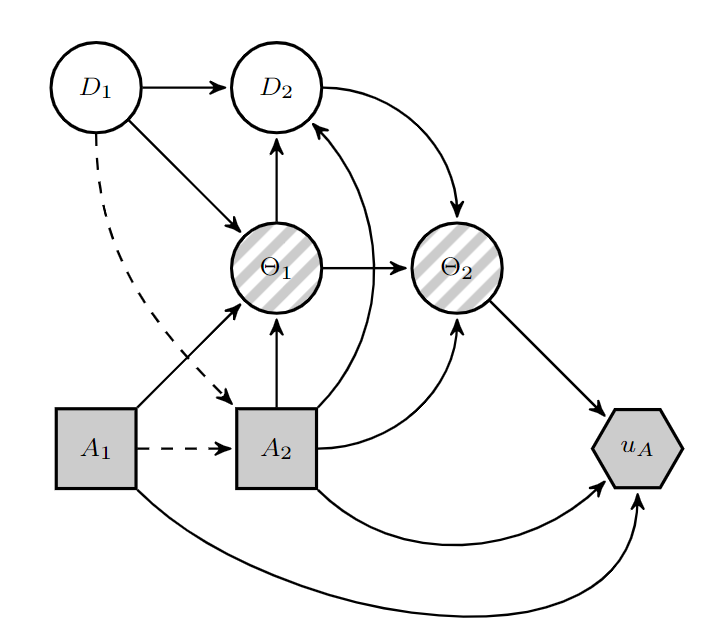} 
\caption{}
\end{subfigure}
\caption{\small (a) BAIDS for DC case study, (b) D's influence 
 diagram, (c) A's influence diagram.}
\label{fig:dismis_ids}
\end{figure}

Our case refers to foreign disinformation interference. We simplify assumptions to better illustrate the proposed concepts 
  although the case is sufficiently complex to reflect the required modeling steps. %
 Figure \ref{fig:dismis_ids}a refers to a scenario with two countries: an attacking one  (\emph{A}) that may attempt to interfere with a defending nation ($\emph{D}$) through a disinformation campaign (DC) during a sensitive moment such as a pandemic.
   \emph{A} seeks to disrupt  $\emph{D}$ through a DC  encouraging $\emph{D}$'s citizens to adopt behaviors leading to infections, increasing $D$'s hospital visits and healthcare costs.
  Even more, if a significant portion of $D$'s population becomes infected, this could overwhelm its health system, potentially requiring the transfer of 
infected citizens to private hospitals, incurring in additional costs. $D$ aims to allocate resources $d_1 \in D_1$ to proactively mitigate attacks by $A$,  say by supporting disinformation 
  debunking research projects.  %
Simultaneously, $\emph{A}$ decides its investment level $a_1 \in A_1$ to
enhance its disinformation capabilities, such as %
   creating more sophisticated bots to disseminate it through social networks \citep{antenore2023comparative}. Subsequently, $\emph{A}$ determines the attack intensity 
$a_2 \in A_2$, for instance, by selecting the channels to propagate the disinformation %
 and by targeting users more likely to spread it %
  across these channels. Suppose $\emph{A}$ obtains information about $\emph{D}$'s investment $d_1$ through its intelligence services. %
   $A$'s investment $a_1$ and attack intensity $a_2$, and 
    $D$'s preparedness $d_1$ collectively determine the extent to which $D$ recognizes that it is being attacked ($\theta_1 \in \Theta_1$).
This understanding enables $\emph{D}$ to implement better reactive measures $d_2 \in D_2$, such as launching advertising campaigns to counter the disinformation spread. 
Actions $d_2$ and $a_2$, taken respectively by $\emph{D}$ and $\emph{A}$ together with $\theta_1$,
influence the number of individuals $\theta_2 \in \Theta_2$ affected by the campaign who, as a result, are more likely to become infected.
This number, combined with the resources allocated at $d_1$ and $d_2$, defines $\emph{D}$'s utility, $u_D(d_1, d_2, \theta_2)$. Similarly, $\emph{A}$'s utility is $u_A(a_1, a_2, \theta_2)$. $\hfill \triangle$

\noindent Figures \ref{fig:dismis_ids}b and \ref{fig:dismis_ids}c 
respectively reflect IDs for 
the Defender and Attacker problems.

\subsection{Methodology motivation}
\label{sec:mot_method}

   Let us first solve $D$'s problem conceptually. Assume for now that we have available the required utilities %
 and probabilities. %
 Then, we treat $D$'s ID as follows, using \cite{Shachter1986} reduction operations.

\begin{itemize}[noitemsep,label={},leftmargin=0pt]
 \item $\mathcal{D}_1$: Remove chance node $\Theta_2$, computing $\psi_D(d_1, d_2, \theta_1, a_2)$=$\int u_D(d_1,d_2,\theta_2)p_D(\theta_2|d_2,a_2,\theta_1)d\theta_2$.
    \item  $\mathcal{D}_2$: Remove decision node $D_2$, computing the optimal  $d_2^{*}(d_1,a_2,\theta_1) = \argmax_{d_2}\psi_D(d_1, d_2,\theta_1, a_2)$ and storing the optimal expected utilities $\psi_D(d_1,a_2,\theta_1) = \max_{d_2}\psi_D(d_1,d_2,\theta_1,a_2)$.
    \item $\mathcal{D}_3$: Remove chance node $\Theta_1$, computing $\psi_D(d_1,a_1,a_2) = \int \psi_D (d_1,a_2, \theta_1)p_D(\theta_1|d_1,a_1,a_2)\dd\theta_1 $. 
    $\mathcal{D}_4$: Remove chance node $A_2$, computing $\psi_D(d_1,a_1) = \int \psi_D(d_1,a_1,a_2)p_D(a_2|d_1,a_1)\dd a_2$. 
   \item  $\mathcal{D}_5$: Remove chance node $A_1$, computing $\psi_D(d_1) = \int \psi_D(d_1,a_1)p_D(a_1)\dd a_1$. %
     \item $\mathcal{D}_6$: Remove decision node $D_1$, obtaining the 
     optimal first stage decision $d_1^{*} = \argmax_{d_1} \psi_D(d_1)$.
\end{itemize}

\noindent Then, the optimal strategy for $D$ would consist of selecting $d^{*}_1$ at node $D_1$, and  %
 $d_2^{*}(d_1^{*},a_2,\theta_1)$ at node $D_2$, given the $a_2$ and $\theta_1$ observed. Note that, in principle, to implement steps $\mathcal{D}_3$-$\mathcal{D}_6$, it is not actually necessary to store the optimal action $d_2^{*}(d_1,a_2, \theta_1)$,
   but just the 
 optimal value $\psi_D(d_1,a_2,\theta_1)$, from which we  
  can subsequently reconstruct $d_2^{*}(d_1^{*},a_2, \theta_1)$. This point is particularly relevant in continuous domains, as it allows for a reduction in storage requirements.  %
Importantly, assuming with no loss of generality, that the utility function $u_D(d_1,d_2,\theta_2)$ is positive, the previous problem may be solved by applying two APS steps:

\begin{itemize}[noitemsep,label={},leftmargin=0pt]
\item {\em DAPS1.} Aggregate steps $\mathcal{D}_1$-$\mathcal{D}_2$, corresponding to $d_2^{*}(d_1,a_2,\theta_1)$=$\argmax_{d_2} \int u_D(d_1,d_2,\theta_2) p_D(\theta_2|\\d_2,a_2,\theta_1)\dd \theta_2$. 
  For this, define the AD $ \pi(d_1,d_2,\theta_2|a_2,\theta_1)\propto u_D(d_1, d_2,\theta_2) p_D(\theta_2|d_2,a_2,\theta_1)$. 
    By sampling from this distribution and consistently estimating $\text{mode}\big(\pi(d_1|a_2,\theta_1)\big)$, we approximate $d_2^{*}(d_1,a_2,\theta_1)$ and store $\widehat{\psi}_D(d_1,a_2,\theta_1) = \pi(d_1, d_2^{*}(d_1,a_2,\theta_1),\theta_2|a_2,\theta_1)$, which is proportional to $\psi_D(d_1,a_2,\theta_1)$.
\item {\em DAPS2.} Aggregate steps $\mathcal{D}_3$-$\mathcal{D}_6$ corresponding to $ d_1^{*} = \argmax_{d_1}   \iiint \widehat{\psi_D}(d_1,a_2,\theta_1) p_D(\theta_1|d_1,a_1,\\a_2) p_D(a_2|d_1,a_1)p_D(a_1)\dd \theta_1 \dd a_2 \dd a_1.$  For this, define the AD 
 \begin{align}   
 \label{eq:DAPS2ad}
\pi(d_1,a_2,\theta_1,a_1) \propto \widehat{\psi}_D(d_1,a_2,\theta_1)p_D(\theta_1|d_1,a_1,a_2)p_D(a_2|d_1,a_1)p_D(a_1). 
\end{align}
    By sampling from it and estimating $\mode(\pi(d_1))$
     consistently, we approximate $d_1^{*}$.
\end{itemize}
Of $D$'s required utilities and probabilities,
  $p_D(a_2|d_1,a_1)$ and $p_D(a_1)$ demand strategic thinking about $A$'s problem. We could solve this should we have the Attacker's utility function, %
  and probabilities.
  However, as argued before, typically $D$ will not be able to obtain such elements precisely. Suppose we represent the corresponding uncertainty with a distribution of random utilities and 
   probabilities $F\sim (U_A(a_1,a_2,\theta_2), P_A(\theta_2|\theta_1,d_2,  a_2), P_A(d_2|d_1,\theta_1,  a_2), P_A(\theta_1|d_1, a_1,a_2), \\P_A(d_1))$ to solve $A$'s problem as follows, where
random expected utilities and random optimal alternatives are 
introduced.
\begin{itemize}[noitemsep,label={},leftmargin=0pt]
\item $\mathcal{A}_1$: Remove $\Theta_2$, computing   $\Psi_A(a_1,a_2,\theta_1,d_2) = \int U_A(a_1,a_2,\theta_2)P_{A}(\theta_2| \theta_1,d_2,a_2) \dd \theta_2$.
\item $\mathcal{A}_2$: Remove $D_2$, computing  $\Psi_A(d_1,\theta_1, a_1, a_2) =  \int \Psi_A(a_1,a_2, \theta_1, d_2)P_A(d_2|d_1,\theta_1,a_2) \dd d_2$. 
\item $\mathcal{A}_3$: Remove $\Theta_1$, computing $\Psi_A(d_1,a_1, a_2) = \int \Psi_A(d_1,\theta_1,a_1,a_2)P_A(\theta _1|d_1,a_1,a_2)\dd{\theta_1}$. 
\item  $\mathcal{A}_4$: Remove $A_2$, computing $ A_2^{*}(d_1,a_1) = \argmax_{a_2} \Psi_A(d_1,a_1,a_2)$
      and storing $\Psi_A(d_1,a_1)$ =$\\\Psi_A(d_1,a_1, A_2^{*}(d_1,a_1))$.
 \item  $\mathcal{A}_5$: Remove $D_1$, computing   $\Psi(a_1) = \int \Psi_A(d_1,a_1)P_A(d_1)\dd d_1$. 
\item  $\mathcal{A}_6$: Remove $A_1$, finding  $A_1^{*} = \argmax_{a_1}\Psi_A(a_1)$.
\end{itemize}

\noindent Through $\mathcal{A}_1$-$\mathcal{A}_6$, we are able to obtain samples from $A_2^{*}(d_1,a_1)$ and $A_1^{*}$ that, respectively, correspond to samples from $p_D(a_2|d_1,a_1)$ and $p_D(a_1)$, the missing ingredients to implement DAPS1-DAPS2. As in $D$'s problem, we only require the random optimal value $\Psi_A(d_1,a_1) = \Psi_A(d_1,a_1, A_2^{*}(d_1,a_1))$  to implement $\mathcal{A}_5$-$\mathcal{A}_6$, which will be relevant when addressing nodes with continuous domains. However, in contrast to the Defender's problem, it is necessary to compute $A_2^{*}(d_1,a_1)$ (or at least approximate it in continuous problems) as it will be used in DAPS2. %
Again, APS allows us to implement the $\mathcal{A}_1$-$\mathcal{A}_6$ operations in a principled manner, through
\begin{itemize}[noitemsep,label={},leftmargin=0pt]
\item {\em AAPS1.} Aggregates steps $\mathcal{A}_1-\mathcal{A}_4$ defining
$A_2^{*}(d_1,a_1) = {\argmax_{a_2} \iiint}  U_A(a_1,a_2,\theta_2)  P_A(\theta_2|\theta_1,d_2,\\a_2)P_A(d_2|d_1,\theta_1,a_2)P_A(\theta_1|d_1,a_1,a_2)\dd \theta_2 \dd d_2 \dd \theta_1.$ 
Consider the  RAD  
$ \Pi(a_1,a_2,\theta_2,d_2,\theta_1|d_1,a_1)\propto U_A(a_1,a_2,\theta_2) P_A(\theta_2|\theta_1, d_2, a_2) P_A(d_2|d_1,\theta_1,a_2)P_A(\theta_1|d_1,a_1,a_2)$.
By sampling from it, which entails previously sampling from each of the aforementioned distributions, we can then estimate $\text{mode}(\Pi(a_2|d_1,a_1))$ consistently, therefore estimating $A_2^{*}(d_1,a_1)$.
\item {\em AAPS2.} Aggregates steps $\mathcal{A}_5-\mathcal{A}_6$. We solve for
\begin{equation*}
A_1^{*} =  \argmax_{a_1} \int \Psi_A(d_1,a_1)P_A(d_1)\dd d_1.
\end{equation*}
Consider the RAD $\Pi(a_1,d_1) \propto \Psi_A(d_1,a_1)P_A(d_1).$ By sampling from it, which entails sampling from $\Psi_A(d_1,a_1)$ and $P_A(d_1)$, we consistently estimate $\text{mode}(\Pi(a_1))$ to approximate $A_1^{*}$.
\end{itemize}

\noindent   In conclusion, by solving two APS for the Defender (DAPS1 and DAPS2) and two random APS (RAPS) for the attacker (AAPS1 and AAPS2) we are conceptually capable of solving the problem. 

Importantly, the approach is actually general. Before discussing this, let us analyse two final examples that introduce additional computational issues not covered in the case. 

\begin{figure}[h!]
  \centering
  \includegraphics[width=0.45\linewidth]{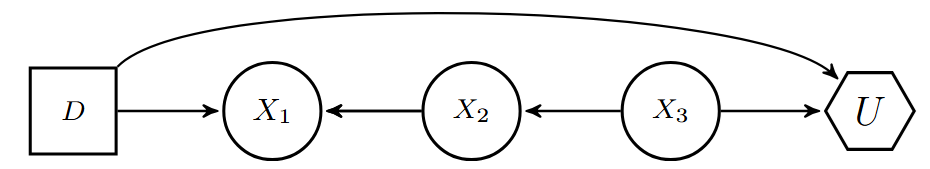} 
  \caption{\small ID for problem where all nodes affected by arc inversion are reduced before node $D$.} \label{fig:example_prop}
\end{figure}

\paragraph{Example 1}{Consider $D$'s ID in Figure \ref{fig:example_prop}. Assume we have $u_D(x_3,d)$, $p_D(x_1|d,x_2)$, $p_D(x_2|x_3)$ and $p_D(x_3)$ available.} Appealing to  Shachter's (1986) algorithm, we would go through the steps

\begin{itemize}[noitemsep,label={},leftmargin=0pt]
\item \noindent {\em Invert arc $X_2-X_3$}, with updated probabilities $p_D(x_3|x_2) = p_D(x_2|x_3)p(x_3)/p_D(x_2)$ and $p_D(x_2) = \int p_D(x_2|x_3)p_D(x_3) \dd x_3$.
\item  {\em  Eliminate $X_3$,} obtaining the expected utility $\psi_D(x_2,d) = \int u_D(x_3, d) p_D(x_3|x_2) \dd x_3$. 
\item {\em  Invert arc $X_1$-$X_2$} with updated probabilities $p_D(x_2|x_1,d) = p_D(x_2)p_D(x_1|x_2,d)/p_D(x_1|d)$ and $p_D(x_1|d) = \int p_D(x_1|x_2,d) p_D(x_2)\dd x_2$. 
\item {\em Eliminate $X_2$,} obtaining the expected utility $\psi_D(x_1,d) = \int \psi_D(x_2,d)p_D(x_2|x_1,d)\dd x_2$. 
\item {\em Eliminate $X_1$,} obtaining the expected utility $\psi_D(d) = \int \psi(x_1,d)p(x_1|d) \dd x_1$. 
\item {\em Eliminate $D$,} obtaining the optimal decision $d^{*} = \argmax \psi_D(d)$.
\end{itemize}

\noindent Observe now that, assuming that $u_D(x_3,d)$ is positive, this can be solved in a single step by considering the AD 
 $\, \pi(d,x_1,x_2,x_3) \propto u_D(x_3,d)p_D(x_1|d)p_D(x_2|x_1,d)p_D(x_3|x_2)    \propto  u_D(x_3,d)p_D(x_1|d,\\x_2)p_D(x_2|x_3)p_D(x_3),$  
defined using only the terms originally available.

Something similar would occur in an Attacker's ID. Assume that $A$'s   problem is also described by an ID as that in Figure \ref{fig:example_prop}. If 
$D$ has access to the random probabilities and utilities $U_A(x_3,d)$, $P_A(x_1|d,x_2)$, $P_A(x_2|x_3)$, $P_A(x_3)$, the problem may be solved 
 with the RAD $\Pi(d,x_1,x_2,x_3) \propto  U_A(x_3,d)P_A(x_1|d,x_2)P_A(x_2|x_3)P_A(x_3)$, again defined using only the original terms. $\hfill\triangle$

\begin{figure}[htbp!]
  \centering
  \includegraphics[width=0.2\linewidth]{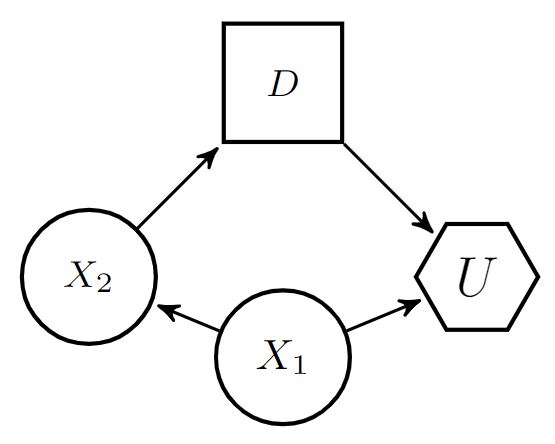} 
  \caption{\small ID for problem where not all nodes affected by arc inversion are reduced before node $D$.} \label{fig:example_prop_2}
\end{figure}

\paragraph{Example 2.} Consider the Defender problem in Figure \ref{fig:example_prop_2}. We have available $u_D(d,x_1)$, $p_D(x_1)$ and $p_D(x_2|x_1)$.
 The steps followed to solve the problem are
 
\begin{itemize}[noitemsep,label={},leftmargin=0pt]
\item {\em Invert arc $X_1\rightarrow X_2$}, computing $ p_D(x_2)=\int p_D(x_2|x_1)p_D(x_1)\dd x_1$ and $p_D(x_1|x_2) =p_D(x_2|x_1)\\p_D(x_1)$$/p_D(x_2)$.
\item {\em Eliminate $X_1$}, computing $\psi_D (x_2,d) = \int u_D(d,x_1)p_D(x_1|x_2)\dd x_1$.
\item {\em Eliminate $D$}, computing $d^{*}(x_2) = \argmax_{d} \psi_D(x_2,d)$.
\end{itemize}

\noindent Observe that, assuming $p_D(x_2)>0$,
\begin{equation*}
 \psi_D(x_2,d) = \int u_D(x_1,d)\frac{p_D(x_2|x_1)p_D(x_1)}{p_D(x_2)}\dd x_1\propto\int u_D(x_1,d)p_D(x_2|x_1)p_D(x_1)\dd x_1.
\end{equation*}

\noindent  Define the AD 
$\,\, \pi_D(x_1, d | x_2) \propto u_D(x_1,d)p_D(x_2|x_1)p_D(x_1) $
and use the fact that $d^{*}(x_2) = \\\text{mode}_{d}(\pi_D(x_1, d | x_2 ))$, to solve the problem with one APS, again defined using only the original terms in the ID. %
$\hfill\triangle$
\vspace{0.1in} 

\noindent Observe that in both examples we were able to define the AD and RAD models in terms of ingredients available 
in the original influence diagrams. In
  the first one, all nodes affected by arc inversions were reduced before node $D$ and, therefore, the marginal terms cancelled. In turn, 
    in the second one, some nodes affected by arc inversions were not reduced before node $D$ but we could ignore the corresponding marginal in the denominator, being a positive constant multiplying the objective function of a maximization problem, therefore not affecting the maximum to be found.

\subsection{General computational scheme}

\citet{BAIDS} introduced a scheme to deal with discrete BAIDs adapting ARA methods \citep{banks16} and classic ID reductions from \citet{Shachter1986}. However,
such reductions were discrete and solvable analytically. Inspired by the reasoning in Sections \ref{sec:n_stage_parallel} and \ref{sec:mot_method} we replace here such reductions
in batches with APS schemes and take advantage of the fact that, for expected utility maximization, we do not need to take into account the denominators in Bayes' formula, as they are positive constants. In this way, we are able to efficiently handle BAIDs with continuous decision and chance nodes providing the first general solution to \cite{Shachter1986}'s query.

Assume that the security game is represented by a proper BAID 
 \citep{banks16}, i.e., an acyclic directed graph over decision, chance and utility nodes, where some chance nodes can be shared by both agents, such that from each decision maker's perspective, the corresponding IDs are proper \citep{Shachter1986}. 
Among other things, this implies that each agent has a directed path connecting their decision nodes, culminating in their respective utility node. We refer to these paths as the \textit{Defender Decision Path}, \(DDP = \{D_1, D_2, \ldots, D_n\}\), and the \textit{Attacker Decision Path}, \(ADP = \{A_1, A_2, \ldots, A_m\}\), where the Defender has \(n\) decisions to make and the Attacker, \(m\) decisions. Besides, the proper BAID definition implies that if two decisions are simultaneous, in the sense of Section \ref{sec:sec_template}, there is no directed path between them. 

Drawing on the case and Section 2, let us introduce two core BAID operations that serve to solve general security problems. As proved in the Appendix, they require the Defender's utility function $u_D$ and, a.s., the random utility function $U_A$ to be positive. 
The first one, DAPS reduction, eliminates the last decision node in the DDP, together with the chance nodes to be eliminated before it, through an APS and serves to find the corresponding Defender's optimal decision, given the information available when making such decision.

\begin{proposition}
\label{prop:DAPS_red}
{\normalfont \textbf{(Defender APS reduction, \texttt{DAPS\_red}).}} Consider the Defender's ID associated to a proper BAID.
  Let $D$ be the last node in its DDP. Let $\mathcal{X}_c(D) = \{X_1, \ldots, X_{n_D}\}$ be the chance nodes to be eliminated before $D$  (possibly including several arc inversions). Then, we can reduce the nodes $(D, \mathcal{X}_c(D))$ using APS with the value node inheriting the antecessors of $\mathcal{X}_c(D)$  not eliminated in this process (and storing the optimal value function).
\end{proposition}

\noindent The second one, AAPS reduction, operates in the Attacker ID. It eliminates the last decision node in the ADP, together with some chance nodes, and serves to find $D$'s distribution over the corresponding Attacker's decisions.

\begin{proposition}
\label{prop:AAPS_red}
{\normalfont \textbf{(Attacker APS reduction, \texttt{AAPS\_red}).}} Consider the Attacker's ID associated to a proper BAID.
  Let $A$ be the last node in its ADP. Let $\mathcal{X}_c(A) = \{X_1, \ldots, X_{n_A}\}$ be the chance nodes to be eliminated before $A$  (possibly including several arc inversions). Then, we can reduce the nodes $(A, \mathcal{X}_c(A))$ using RAPS with the value node inheriting the antecessors of $\mathcal{X}_c(A)$  not eliminated in this process. The random optimal conditional decisions are recorded to predict the Attacker's decisions.
\end{proposition}
Note that \texttt{DAPS\_red} exclusively needs the probability distributions originally available to $D$, allowing efficient sampling from these distributions,  avoiding costly arc inversions, and 
greatly facilitating the computation of the optimal decision. Similarly, 
 \texttt{AAPS\_red} requires only the random probability distribution initially assigned by $D$ to $A$, thus allowing for efficient execution of the operation and, actually, not requiring explicit arc inversions.

Based on both results we provide Algorithm 2 to solve general proper BAIDs regardless of the number of Defender's and Attacker's decision nodes and their temporal interrelations. We use the notation $d^{*}(\cdot )$ to indicate the arguments $(\cdot )$ upon which the optimal decision $d^{*}$ at a node depends; similarly, $A^{*}(\cdot)$ refers to the variables upon which the random optimal decision $A^{*}$ at a given node depends. Algorithm \ref{alg:ARA_APS_SIM_A} outputs the optimal decision $d_i^{*}(\cdot)$ for $D$ at each of her decision nodes $D_i$ and, as useful complementary information, the forecast $A_j^{*}(\cdot)$ of $A$'s actions at each of his decision nodes $A_j$, taking as input the BAID $B$ with $D$'s distributions and $A$'s random distributions.

\vspace{0.3cm}
\SetKwFunction{Fatk}{solve\_BAID\_ARA\_APS}
{\small
\begin{algorithm}[H]
\linespread{0.7}\selectfont
\; 
\SetKwProg{Fn}{function}{:}{}
\Fn{\Fatk{$B$}}{
Define Attacker ID and Identify ADP\;
Define Defender ID and Identify DDP\;
\While{DDP $\neq $$\emptyset$}
{Find last decision node $D_i$ in DDP \;
\If{\textnormal{all} $p_D$ \textnormal{for} \texttt{DAPS\_red} \textnormal{of} $D_i$ \textnormal{available}}{
Apply \texttt{DAPS\_red} and store optimal decision $d^{*}_i(\cdot)$\;
Update Attacker's beliefs\;
Eliminate $D_i$ from DDP}
\Else{
Find last decision node $A_j$ in ADP \;
Apply \texttt{AAPS\_red} and obtain optimal random decision $A_j^{*}(\cdot)$\;
Update Defender's beliefs \;
Eliminate $A_j$ from ADP
}
}
\textbf{Return: $\{d^{*}_1(\cdot), \ldots, d^{*}_n(\cdot),A^{*}_1(\cdot), \ldots, A^{*}_m(\cdot)\}$}}
\caption{\small General algorithm to solve BAIDs using ARA and APS.}\label{alg:ARA_APS_SIM_A}
\end{algorithm}}
\vspace{0.3cm}

\noindent Algorithm \ref{alg:ARA_APS_SIM_A} implements $n$ DAPS reductions
 and $m$  AAPS reductions to provide the optimal policy, as  \ref{sec:app_proof_pro} proves. 

\begin{proposition}
\label{prop:gen_algo}
In a proper BAID, Algorithm \ref{alg:ARA_APS_SIM_A} provides the optimal policy for the Defender.
\end{proposition}

\section{Computational issues}
\label{sec:computational_issues}

Algorithm \ref{alg:ARA_APS_SIM_A} utilizes backward dynamic programming and, implicitly, probabilistic inversion. It is directly applicable only in security games with discrete and low-cardinality decision and uncertainty spaces. To illustrate this, consider the DAPS1 step when solving the problem 
  in Figure \ref{fig:dismis_ids}b. To find the optimal $d_1^*$ using APS, we sample from the AD in (\ref{eq:DAPS2ad}). This  is typically performed with MH, creating a Markov chain on ($d_1, a_2, \theta_1$, $a_1$) by proposing candidates for these variables and updating them based on an acceptance probability that depends, among other factors, on $\psi_D(d_1,a_2,\theta_1)$ evaluated at both the old and new candidate values. If the decision and outcome spaces have sufficiently low cardinality, it is practical to solve DAPS2 to compute $\psi_D $ at every feasible $(d_1,a_2,\theta_1)$. Similarly, empirical estimates of $p_D(a_2 | d_1, a_1)$ can be built for all feasible $(d_1, a_1)$.
 However, this approach is not applicable in continuous decision spaces and, in the discrete case, can lead to a combinatorial explosion as the cardinality of decision spaces and the number of stages increase. In such scenarios, it is necessary to develop 
  models that can effectively approximate future optimal decisions and predict attack distributions based on a few cases. This section outlines computational details necessary to efficiently apply Algorithm \ref{alg:ARA_APS_SIM_A} under these circumstances,  illustrated in Section \ref{sec:casestudy}.

\subsection{Approximating Defender's optimal value functions}
\label{sec:storeD}

In continuous problems or problems with high cardinality decision and/or uncertainty spaces, computing optimal values for the defender, such as $\psi_D(d_1,a_2,\theta_1) = \max_{d_2} \psi_D(d_1, d_2, a_2, \theta_1)$ for every $(d_1,a_2,\theta_1)$ possible becomes infeasible. However, we can approximate optimal value functions with a statistical metamodel \citep{law2007simulation} to be 
propagated backwards to earlier decision stages. To illustrate this, let us discuss   DAPS1 in the disinformation war case, though the arguments are general.

To develop the metamodel, collect first data from simulations that represent various scenarios within the $(d_1, a_2, \theta_1)$ spaces, 
  covering a broad range of values through a grid 
 $(d_1, a_2, \theta_1)_{j=1}^J$ to ensure metamodel robustness. For each $j$, the optimal $d_2$ is computed by invoking DAPS1, and $\psi_D(d_1,a_2,\theta_1)_j$ is estimated via MC. Denote the data by $\mathcal{D} = \{(d_1, a_2, \theta_1)_j,  \psi_D(d_1, a_2, \theta_1)_j \}_{j=1}^J$. %
Next, a regression metamodel $\widehat{\psi}_{D,\gamma}(d_1, a_2, \theta_1)$ with covariates $(d_1, a_2, \theta_1)$ and parameterized by $\gamma$ is trained to predict $\psi_D(d_1, a_2, \theta_1)$ using  $\mathcal{D}$. The selection of an appropriate statistical modeling technique is crucial, with neural networks (NN) \citep{gallego2022current} as prime example. The metamodel's performance is assessed using metrics such as mean squared error, and evaluated on a hold-out dataset;  model parameters are tuned mainly through cross-validation. %
    This process of model development and refinement iterates 
      to achieve better performance.
Once a sufficiently good metamodel is available, it is integrated within the backward induction process. In our particular case, 
when simulating from the AD in (\ref{eq:DAPS2ad}) using MCMC to solve DAPS2, the metamodel $\widehat{\psi}_{D, \gamma}(d_1,a_2,\theta_1)$ will be used instead of the original optimal value function, allowing us to evaluate it for every possible candidate value. However, we still need to be able to sample from $p_D(a_2|d_1,a_1)$ for every possible $d_1, a_1$. We explain how to build an approximate sampler next.

%
%

%

\subsection{Approximately sampling from random optimal attacks}
\label{sec:appr_A_PSI_attack}

Again, in continuous problems or problems with high cardinality decision and/or uncertainty spaces, sampling from attack distributions such as $p_D(a_2 | d_1, a_1)$ for every possible $(d_1,a_1)$ or, equivalently, computing random optimal values for the Attacker, such as $\Psi_A(d_1,a_1) = \max_{a_2}\Psi(d_1,a_1,a_2)$ becomes inefficient, even infeasible. 
  A feasible approach is to approximate the attack distribution and  random optimal value function with a statistical metamodel, to be propagated backward to earlier stages. To illustrate this, let us discuss  AAPS1 in the case though, again, the arguments are general. 

As before, we collect data from simulations for a broad range of values for $d_1, a_1$ to ensure robustness, through a grid $(d_1, a_1)_{j=1}^J$, where $J$ represents the number of data points. For each value in the grid, we produce a sample from the conditional probability $p_D(a_2 | d_1, a_1)$ using AAPS1. Let $\mathcal{D} = \left\lbrace (d_1, a_1)_j, \{a_{2}\}_{k=1}^K \right\rbrace_{j=1}^J$ denote the data, with $K$ the number of samples for each point in the grid. Once $\mathcal{D}$ has been obtained, $\{\Psi_A(d_1, a_1)\}_{k=1}^K$ is computed using MC simulation, resulting in the dataset $\mathcal{D}' = \{(d_1, a_1)_j, \big(\{\Psi_A(d_1, a_1)\}^K_{k=1}\big)_j\}_{j=1}^J$. Next, we choose appropriate statistical models $\widehat{p}_{D}(a_2 | d_1, a_1)$ and $\widehat{\Psi}_A(d_1, a_1)$ to approximate the attack distribution and the random optimal value function, respectively. Methods such as kernel density estimation, Gaussian processes or neural networks \citep{bishop2006pattern} can be used as basic models. Note that if the attack decision space or random value function are bounded, a transformation may be necessary before modeling. Model performance is evaluated using likelihood-based measures, assessed on a hold-out dataset. Parameters are tuned through cross-validation to improve accuracy, with possibly several iterations
   of model development and refinement as necessary to achieve better performance.
Once the models are developed, they are integrated within the backward induction process. Specifically,  $\widehat{p}_{D}(a_2 | d_1, a_1)$ is used in place of the original distribution to sample efficiently from the approximated attack distribution for every 
   $(d_1, a_1)$, which will be used to solve DAPS2. Similarly, $\widehat{\Psi}_A(d_1, a_1)$ will be used instead of the original random optimal value function when using MCMC to solve AAPS2.

%

\subsection{Availability of probabilities}

 Algorithm \ref{alg:ARA_APS_SIM_A} includes two {\em update beliefs} statements.  First, the \texttt{if} condition demands whether all $p_D$ are available to implement \texttt{DAPS\_red} for the last decision node $D_i$ in $DPP$. To check this, we qualitatively apply Shachter's algorithm to find $\mathcal{X}_c(D_i)$. If none of the nodes in this set corresponds to Attacker's uncertainties that have not been treated before, then declare all $p_D$ as available. Otherwise, we move to the {\bf else} block and reduce the required
$A_j$ nodes in the ADP inverse sequence order until
deducing the corresponding attack distributions based on Section 4.2
and updating the corresponding $p_D(a_j|.)$
distributions.

The second beliefs update refers to assessing the Attacker's beliefs for the purpose of later iterations. 
 Originally we set up such distributions as non-informative
  ones,
 given the lack of knowledge. At later stages, 
    as we find the optimal decisions $d_i^{*}(\cdot)$, 
    we can reassess the $p_A(d_i|.)$ distributions as
    centered around $d_i^{*}(\cdot )$, with some uncertainty around them.

\subsection{Enhancing the efficiency of APS} \label{sec:enhance_APS}
In all $D$ and $A$ stages,  we increase the efficiency of APS by sampling from a power transformation of the corresponding marginal AD or RAD, see \cite{muller2004optimal}: we substitute the marginal augmented distribution $\pi(\cdot)$ by a power transformation $\pi^h(\cdot)$ where $h$ is an augmentation parameter. This transformation is more peaked around the mode, facilitating mode identification. We focus on the Defender's problem in Algorithm \ref{alg:ARA_APS_framework} (\texttt{DAPS}) to illustrate how to sample from $\pi^h(\cdot)$. Specifically, in Algorithm  \ref{alg:ARA_APS_framework}, sampling from $\pi^h(d)$ can be performed by drawing $h$ copies of $\tilde{a}$ and $\tilde{\theta}$, instead of just one, and modifying the acceptance probability at the $i$-th iteration by  $\min\left\{1,\frac{g(d^{(i-1)}|\tilde{d})}{g(\tilde{d}|d^{(i-i)})}\prod_{t=1}^{h}\frac{u_D(\tilde{d},\tilde{\theta_t})}{{u_D(d^{(i-1)},\theta_t^{(i-1)})}}\right\}$.

\section{Case study}
\label{sec:casestudy}

\subsection{Basic modeling assumptions}

 Let us illustrate the methodology with 
a numerical instance of the disinformation war case in Figure \ref{fig:dismis_ids}. The supplementary materials (SM) detail $D$'s 
parametric assessments concerning both her decision-making problem and her perspective on how $A$ would address his. Table \ref{tab:ex_summary} summarises the nodes in the BAID and their ranges.  Feasible values for $D_1$ ($A_1$) are viewed as 
  the proportion of the available budget that $D$ ($A$) aims to allocate. For example, $d_1 = 0$ indicates a minimum investment in $D$'s defense program; 
  similarly, $A_2$ corresponds to the intensity with which $A$ aims the attack to occur, with $a_2=0$ meaning that he decides not to attack; higher values increase the number of individuals exposed to the campaign and enhance its effectiveness. $d_2$ represents the proportion of available budget for $D_2$ utilised by $D$ for reactive countermeasures.
$\theta_1$ represents the degree with which the Defender recognizes the disinformation campaign, allowing resources invested in $D_2$ to be more effective: higher $\theta_1$ suggests a better detected attack. 
$\Theta_2 \in \{0,1, \ldots, n \}$ represents the number of citizens susceptible of becoming affected by the campaign and, thus, will be likely to be infected. $D$'s utility $u_D$ and $A$'s utility $u_A$
respectively range in $(0, y_D]$ and $(0, y_A]$.

\begin{table}[!h]
\begin{adjustbox}{width=0.70\columnwidth,center}
\centering
\begin{tabular}{c|c|c}
\hline
Node                       & Concept                                                                                                                                                               & Range         \\ \hline
$D_1$                      & \begin{tabular}[c]{@{}c@{}}Resources to proactively  defend  against disinfo\end{tabular}                                             & {[}0,1{]}             \\ \hline
$D_2$ & \begin{tabular}[c]{@{}c@{}} Resources to reactively  protect against disinfo\end{tabular}                                              & [0,1]           \\ \hline
$A_1$                      & \begin{tabular}[c]{@{}c@{}} Resources for more efficient  disinfo campaign\\ \end{tabular} & [0,1]             \\ \hline
$A_2$                      & Intensity of disinfo campaign                                                                                                     & {[}0,1{]}           \\ \hline
$\Theta_1$                      & \begin{tabular}[c]{@{}c@{}}Recognition degree of disinfo campaign\end{tabular}                                         & {[}0,1{]}           \\ \hline
$\Theta_2$                      & \begin{tabular}[c]{@{}c@{}} Individuals  affected by disinfo campaign\end{tabular}                                                         & $\{0,1,\ldots, n\}$ \\ \hline
$u_D$                      & $D$'s utility                                                                                                                                                             & (0,$y_D$]                  \\ \hline
$u_A$                      & $A$'s utility                                                                                                                                                             & (0,$y_A$]                  \\ \hline
\end{tabular}
\end{adjustbox}
\caption{\small Nodes, interpretation, and feasible values for BAID in 
 Figure \ref{fig:dismis_ids}a.}
\label{tab:ex_summary}
\end{table}

%
\subsection{Results}
\label{sec:results}

This section presents the results of the proposed analysis providing $D$'s optimal decisions at nodes $D_1$ and $D_2$. 
Additionally, our approach provides probabilistic predictions of $A$'s actions at nodes $A_1$ and $A_2$,
serving to compute $D$'s optimal decisions but also to raise awareness about likely attacks. In all decision stages, we use the method described in Section~\ref{sec:enhance_APS} to enhance APS efficiency, with augmentation parameters $h$ at various stages provided in the  SM.

Given the continuous nature of the decision and uncertainty spaces, we draw on the approaches discussed in Section \ref{sec:computational_issues}. 
We approximate $ \psi_{D}(d_1, a_2, \theta_1)$ (Section \ref{sec:storeD}), $ p_{D}(a_2 | d_1, a_1)$ and $ \Psi_A(d_1, a_1)$ (Section \ref{sec:appr_A_PSI_attack}) 
with multilayer perceptrons (MLPs) \citep{gallego2022current}. Regarding $\widehat{\psi}_{D}(d_1, a_2, \theta_1)$, the MLP is trained to approximate a scalar function by minimizing mean squared error \citep{bishop2006pattern} using the dataset $\mathcal{D}_{\psi_{{D}_2}} = \{(d_1, a_2, \theta_1)_j,  \psi_D(d_1, a_2, \theta_1)_j \}_{j=1}^J$. For $\widehat{p}_{D}(a_2 | d_1,  a_1)$\big($\widehat{\Psi}_A(d_1, a_1)$\big), the MLP is trained to approximate density functions by minimizing the negative log-likelihood (NLL) of the distribution it parametrizes, evaluated at the samples in the dataset $\mathcal{D}_{{A}_2} = \{(d_1, a_1)_j, \{a_2^*(d_1,a_1)\}_{k=1}^K \}_{j=1}^J$ \big($\mathcal{D}_{{\Psi_{A_2}}} = \{(d_1, a_1)_j, \big(\{\Psi_A(d_1, a_1)\}^K_{k=1}\big)_j\}_{j=1}^J \big)$.

To select the best MLP architecture for each case, and evaluate the performance of the models, we proceed as follows: we split the corresponding dataset $\mathcal{D}_i, i\in\{\psi_{{D}_2}, A_2, \Psi_{A_2}\}$, into 80\% for training and 20\% for testing. The training set is used to select the MLP architecture via 5-fold cross-validation with early stopping, repeating the process 10 times.  Using the selected architecture, we train an MLP on the full training set and evaluate it on the test set to assess performance. Finally, we retrain the MLP with the selected hyperparameters on the entire dataset $\mathcal{D}_i$ to obtain the model to be propagated backwards to earlier decision stages.

For each stage, we present only the finally selected architecture for the corresponding NN.  Details regarding their selection 
and performance 
are provided in the SM. 

\subsubsection{DAPS1. Defender's second stage decision} \label{sec:def_second_stage_decision}

For a grid of $(d_1, a_2, \theta_1)$ values we:
(a) compute the optimal decision $d_2^*(d_1, a_2, \theta_1)$; and (b) approximate $\psi_D(d_1, a_2, \theta_1) = \max_{d_2} \psi_D(d_1, d_2, a_2, \theta_1)$, as in Section \ref{sec:storeD}, 
based on a grid of $J = 9621 = 21^3$ $(d_1, a_2, \theta_1)$ values, obtained by uniformly dividing the sets $D_1$, $A_2$, and $\Theta_1$ with  splits of 0.05. %

\begin{itemize}[noitemsep,label={},leftmargin=0pt]

\item(a) The optimal $d_2^*(d_1, a_2, \theta_1)$ is obtained by invoking DAPS1 for each $(d_1, a_2, \theta_1)$, obtaining $\mathcal{D}_{{D}_2} = \left\lbrace (d_1, a_2, \theta_1)_j, d_2^*(d_1, a_2, \theta_1)_j \right\rbrace_{j=1}^J$.  Figure \ref{fig:colormap_D2_theta2} (Left) displays the mean of \( d_2^*(d_1,a_2,\theta_1) \) over the values of \( d_1 \); Figure \ref{fig:colormap_D2_theta2} (Right) shows the expected number
\( E[\theta_2|d^*_2(d_1,a_2,\theta_1),a_2,\theta_1] \)
of infected individuals, obtained by %
determining \( d_2^*(d_1,a_2,\theta_1) \) and performing a Monte Carlo (MC) simulation for each \(\big((d^*_2(d_1,a_2,\theta_1))_j, (a_2, \theta_1)_j\big)\). Colors in this Figure indicate whether the infected individuals can be accommodated within \( D \)'s public healthcare system (green) or if excess cases must be transferred to the private sector (red), showcasing the protection level that $d^*_2$ provides to $D$'s health system. Observe that \( E[\theta_2 \mid d^*_2(d_1, a_2, \theta_1), a_2, \theta_1] \) could have been computed analytically. However, we chose to use MC simulation to approximate this expectation in order to illustrate how to conduct the analysis in general, regardless of whether the analytical form of the chance node is known.
\begin{figure}[h]
    \centering    \includegraphics[width=0.8\textwidth]{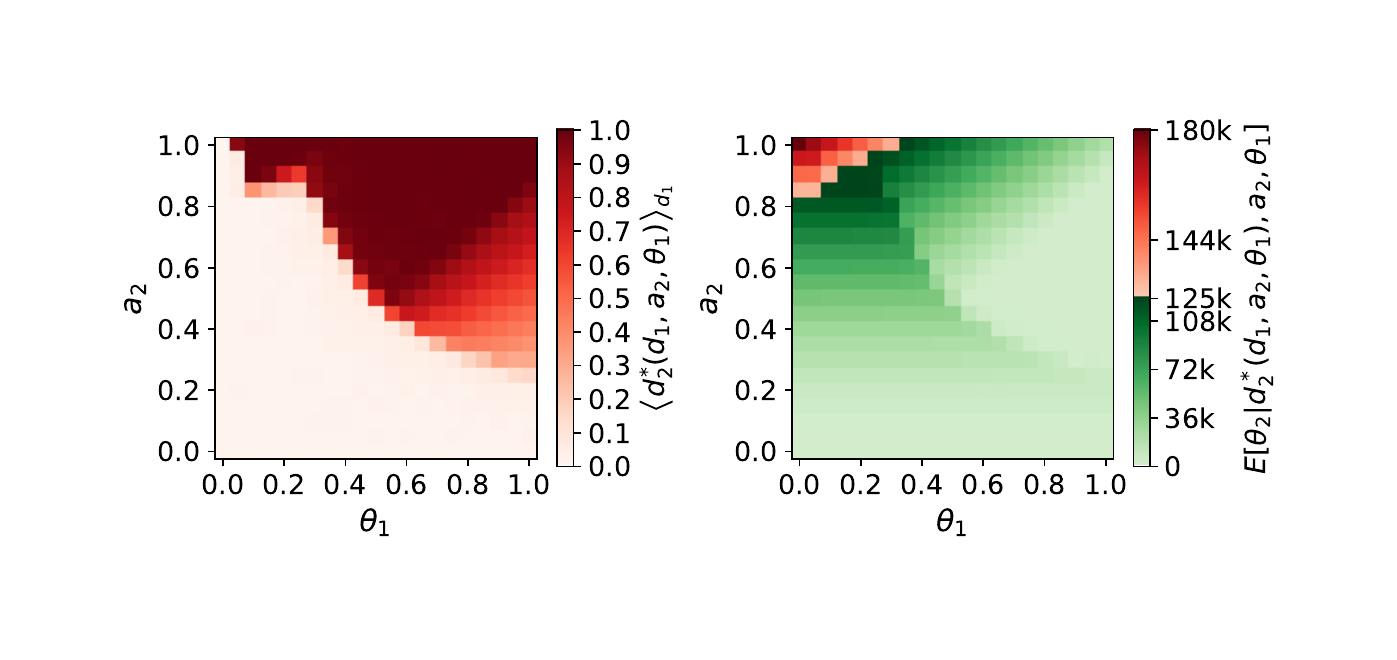}
    \vspace{-1.2cm}
    \caption{\small (Left) Mean $\langle d_2^*(d_1,a_2,\theta_1) \rangle_{d_1}$ of \( d_2^*(d_1,a_2,\theta_1) \) over \( d_1 \). (Right) Expected number \( E[\theta_2 \mid d^*_2(d_1,a_2,\theta_1),a_2,\theta_1] \) of individuals to be infected. Best viewed in color.}
    \label{fig:colormap_D2_theta2}
\end{figure}

  Figure \ref{fig:colormap_D2_theta2} (Left) reveals three distinct behaviors of $d_2^*(d_1,a_2, \theta_1)$. First, practically no defense ($d_2^*(d_1,a_2, \theta_1) \approx 0$, white region) is deployed when recognition ($\theta_1$) is low, except for high intensity attacks ($a_2 \gtrsim
0.9$), as $d_2$ would be ineffective in reducing infections;  
  additionally, no defense is deployed when healthcare costs are lower than the required $d_2$ investment, even at high $\theta_1$, since $D$'s health system can manage the outbreak. Second, an intermediate defense ($0 < d_2^*(d_1,a_2,\theta_1) < 1$, pale to medium red) occurs when an increase in recognition ($\theta_1$) enhances the efficiency of $d_2$, reducing infections with a lower investment, thus preventing both $D$'s under- and overspending.  Finally, full deployment of defenses ($d^*_2(d_1,a_2,\theta_1) \approx 1$, dark red) takes place when recognition ($\theta_1$) is low under a high-intensity attack ($a_2 \gtrsim 0.9$), in an effort to mitigate numerous infections that may have severe consequences, even if the approach is inefficient. This option is also preferred when the investment prevents infected individuals from being transferred to the private sector, as the number of infections remains within the healthcare system's capacity (red transitioning to green in Figure \ref{fig:colormap_D2_theta2} (Right)).

\item(b) For each $\big((d_1, a_2, \theta_1)_j, d_2^*(d_1, a_2, \theta_1)_j)\big)$ in $\mathcal{D}_{{D}_2}$, we compute $\psi_D(d_1, a_2, \theta_1)$ via MC obtaining  $\mathcal{D}_{{\psi}_{D_2}}$. We approximate $\psi_D(d_1, a_2, \theta_1)$ by fitting a MLP with three hidden layers containing 32, 64, and 16 neurons, respectively, and ReLU activation functions. %
  This approximation is used when computing the Attacker’s first-stage decision. %
\end{itemize}

\subsubsection{AAPS1. Prediction of attacker's second stage decision}
\label{sec:att_second_stage_decision}

For a grid of values $(a_1, d_1)$, we approximate
(a) the random optimal decisions $A_2^*(d_1, a_1)$ with density $p_D(a_2 | d_1, a_1)$; and (b) the random expected utility $\Psi_A(d_1, a_1) = \max_{a_2} \Psi_A(d_1, a_1, a_2)$.  
The grid used has $J = 41^2 = 1681$ combinations of $(d_1, a_1)$ generated by uniformly splitting $D_1$ and $A_1$ with splits of 0.025.

\pagebreak
\begin{itemize}[noitemsep,label={},leftmargin=0pt]
\item(a)  Compute $\{a^*_{2}(d_1, a_1)_k\}_{k=1}^{100}$ for each $(d_1, a_1)$  invoking AAPS1, 
thereby simulating the behavior of 100 attackers at each point, leading 
to the dataset ${\large 
\mathcal{D}_{A_2}}$. Figure \ref{fig:A2_results} displays the estimated expected random optimal attack ($\overline{A}_2^*(d_1,a_1)$), the mean of $\{a^*_{2}(d_1, a_1)_k\}_{k=1}^{100}$ for each $(d_1,a_1)$.

 \begin{figure}[h]
    \centering    \includegraphics[width=0.31\textwidth]{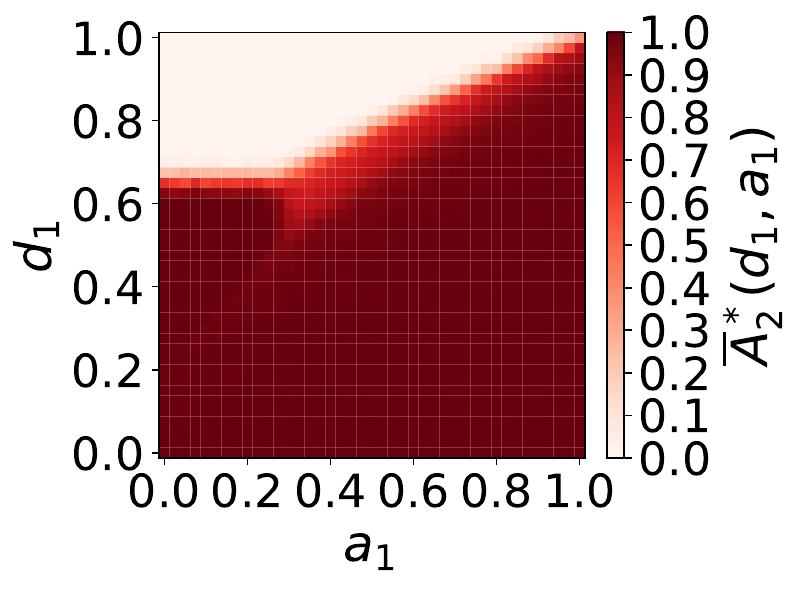}
    \caption{\small  Expected random optimal attack $A_2^*(d_1, a_1)$ in the second stage. Best viewed in color.}
    \label{fig:A2_results}
\end{figure}

  Figure \ref{fig:A2_results} suggests three behaviors for  $\overline{A}_2^*(d_1,a_1)$:  (1) $\overline{A}_2^*(d_1,a_1) \approx 0$ (white region), occurring when $d_1$ is much higher than $a_1$; $A$ observes that $D$ invested significantly more in $d_1$ than he has in  $a_1$, deciding not to attack, as success is unlikely; 
(2) $\overline{A}_2^*(d_1,a_1) \approx 1$ (dark red): when $a_1$ is low, $A$ launches a high-intensity attack as long as $D$'s proactive investment remains moderate ($d_1 \lesssim 0.6$). As $a_1$ increases, $A$ continues to launch high-intensity attacks even at higher $d_1$, as greater investment in $a_1$ enables implementing more impactful attacks ($a_2$) that remain effective despite stronger defensive measures by $D$; finally, (3) when $0 < \overline{A}_2^*(d_1,a_1) < 1$ (moderate red): $A$ launches an intermediate-intensity attack that is sufficiently effective to harm $D$, but refrains from a high-intensity attack, as the cost of a stronger $a_2$ is not justified by its expected impact given the investments $(d_1, a_1)$.

   We approximate $A_2^*(d_1, a_1)$ using the dataset $\mathcal{D}_{A_2}$ with an MLP that outputs  parameters $\{(w_i, \alpha_i, \beta_i)\}_{i=1}^2$ defining a mixture of Beta distributions for each input $(d_1, a_1)$. The MLP consists of three hidden layers with 64 neurons each, using ReLU activation functions in the intermediate layers, a softmax activation for the output neurons corresponding to $\{w_i\}_{i=1}^2$, and a softplus activation for those corresponding to $\{(\alpha_i, \beta_i)\}_{i=1}^2$. The approximation is integrated into the Defender’s first-stage decision process.

\item(b) For each $(d_1, a_1)$ in $\mathcal{D}_{A_2}$, we compute $\Psi_A(d_1, a_1)$ via MC simulation, resulting in $\mathcal{D}_{\Psi_{A_2}}$. 
  To approximate $\Psi_A(d_1,a_1)$, we fit to $\mathcal{D}_{\Psi_{A_2}}$ an MLP that outputs the parameters $\{(w_i, \lambda_i, \alpha_i)\}_{i=1}^2$ characterizing a mixture of Weibull distributions for each $(d_1,a_1)$. Again, the MLP comprises three hidden layers with 64 neurons each, employing ReLU activations in the hidden layers, softmax activation for the output neurons associated with $\{w_i\}_{i=1}^2$, and softplus activation for those corresponding to $\{(\lambda_i, \alpha_i)\}_{i=1}^2$. $\widehat{\Psi}_A(d_1, a_1)$ is used in the Attacker's first stage.

\end{itemize}

%

\subsubsection{ Prediction of attacker's first stage decision} \label{sec:att_1st_dec}

\begin{figure}[!h]
    \centering    \includegraphics[width=0.25\textwidth]{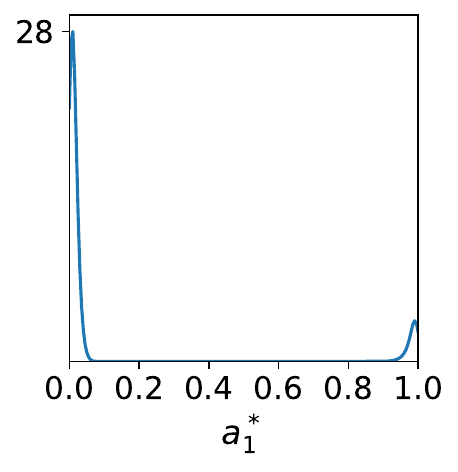}
    \vspace{-0.9em}
    \caption{\small Density of empirical distribution $A_1^*$.}
    \label{fig:a1_vs_moB}
\end{figure}

Using $ \widehat{\Psi}_A(d_1, a_1)$, we invoke AAPS2 to generate 10000 samples of $A_1^*$ and display their density ($p_D(a_1)$) in Figure \ref{fig:a1_vs_moB}.  Observe that, from $D$'s perspective, $A$ adopts two strategies: most likely choosing not to allocate resources to produce a more effective disinformation campaign ($a_1 \approx 0$), or, though much less likely, investing all resources ($a_1 \approx 1$). A mixture of Beta distributions is employed to approximate $A_1^*$, and used in the Defender's first stage. 

\pagebreak
\subsubsection{DAPS2. Defender's first stage decision}\label{sec:def_2_decision}

With the approximations  $\widehat{\psi}_D(d_1, a_2, \theta_1)$, $\widehat{p}_D(a_2 | d_1, a_1)$, and $\widehat{p}_D(a_1)$, finding the Defender's first stage decision involves applying APS %
by invoking DAPS2. 
The optimal decision is $d^*_1 =0.7$, suggesting that $D$'s best strategy involves allocating a moderate proportion of her budget to proactive measures. 

\begin{figure}[h]
    \centering    \includegraphics[width=0.32\textwidth]{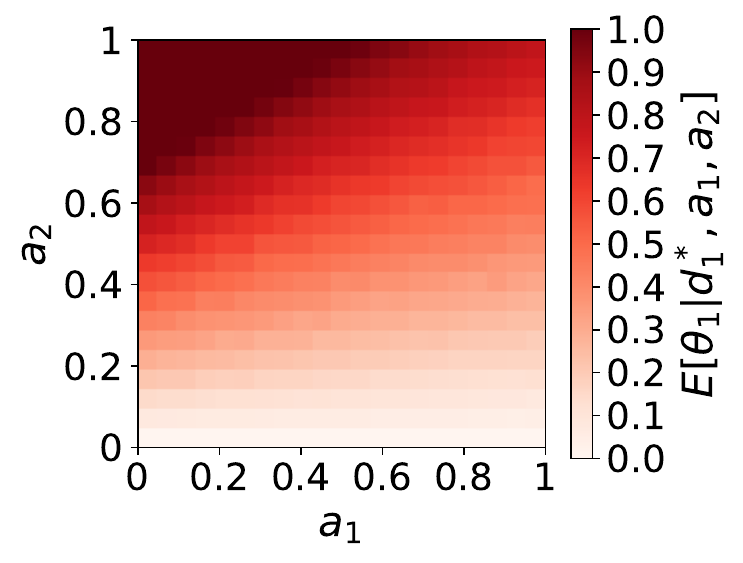}
    \caption{\small Expected degree of recognition 
    $ E[\theta_1 \mid d_1^*, a_1, a_2]$ for $d_1^*$ across different values of $(a_1, a_2)$. Best viewed in color.}
    \label{fig:expected_theta1_d1_star}
\vspace{-0.5cm}
\end{figure}

Let us analyze the implications of $d_1^*$ through Figure \ref{fig:colormap_D2_theta2} (Right), Figure \ref{fig:A2_results} and Figure \ref{fig:expected_theta1_d1_star}, displaying $E[\theta_1|d_1^*, a_1, a_2]$. 
Notice two effects: first, $d_1^*$ acts as a deterrent, leading $A$  to not attack on average when it has previously invested at least a moderate to low amount in $A_1$ ($a_1^* \approx 0.3$), as observed in Figure \ref{fig:A2_results}; second, if $A$ decides to attack, assuming $a_1 > 0.3$, either with a moderate or high-intensity attack \big(Figure \ref{fig:A2_results}\big), $d_1^*$ enables a moderately high degree of recognition $\theta_1$ (Figure \ref{fig:expected_theta1_d1_star}), helping to mitigate $a_2$, as countermeasures $d_2$ will be more efficient, contributing to  reducing the potential harm to $D$’s health system \big(Figure \ref{fig:colormap_D2_theta2}(Right)\big).

\subsubsection{Summary of Defender's optimal policy and 
Attacker's random optimal policy.}
\begin{itemize}[noitemsep,label={},leftmargin=0pt]

\item \textit{Defender's optimal policy.} 
In the first stage, the Defender invests moderately in proactive defense against disinformation ($d_1^* = 0.7$). Given the Attacker's random optimal policies for both decisions ($A_1^*$, $A_2^*(d_1, a_1)$), the Defender faces the following scenarios according to her beliefs: either $a_2 \approx 0$ or $a_2 \approx 1$. Specifically, if $a_2 \approx 0$, the Defender should not implement reactive defenses ($d_2^* \approx 0$); otherwise, if $a_2 \approx 1$, the Defender should use all its resources to protect against the attack ($d_2^* \approx 1$).

\item \textit{Attacker's random optimal policy.} %
In the first stage, the Attacker, most likely according to her beliefs, does not invest in a more effective disinformation campaign ($a_1 \approx 0$); however, in rare cases, the Attacker may commit all available resources to improve its capabilities ($a_1 \approx 1$). In the second stage, if $a_1 \approx 0$, the Attacker will refrain from launching an attack ($a_2 \approx 0$); alternatively, if $a_1 \approx 1$, the Attacker will carry out a maximum-strength attack ($a_2 \approx 1$).

\end{itemize}

\subsection{\textbf{Sensitivity analysis}}
\label{sec:sensi_ana}

We perform sensitivity analysis on two key parameters specified in the SM. 

\subsubsection{Effect of $\omega_{d_2}$ on the Defender's second stage decision}

This parameter determines the effectiveness of the defensive measures $d_2$ and the recognition level $\theta_1$ in countering attack $a_2$. When $\omega_{d_2} > 1$, $d_2$ is more effective than $a_2$,
   neutralizing it even without full recognition (\(\theta \neq 1\)) or maximum investment in $d_2$ $( \neq 1)$;  when $\omega_{d_2} = 1$, full recognition (\(\theta = 1\)) and maximal allocation to $d_2$ ($= 1$) are required to entirely stop $a_2$; finally, when $\omega_{d_2} < 1$, even with full resource allocation and recognition, it may not be feasible to halt the attack spread.

\begin{figure}[!h]
    \centering
    \includegraphics[width=0.94\textwidth]{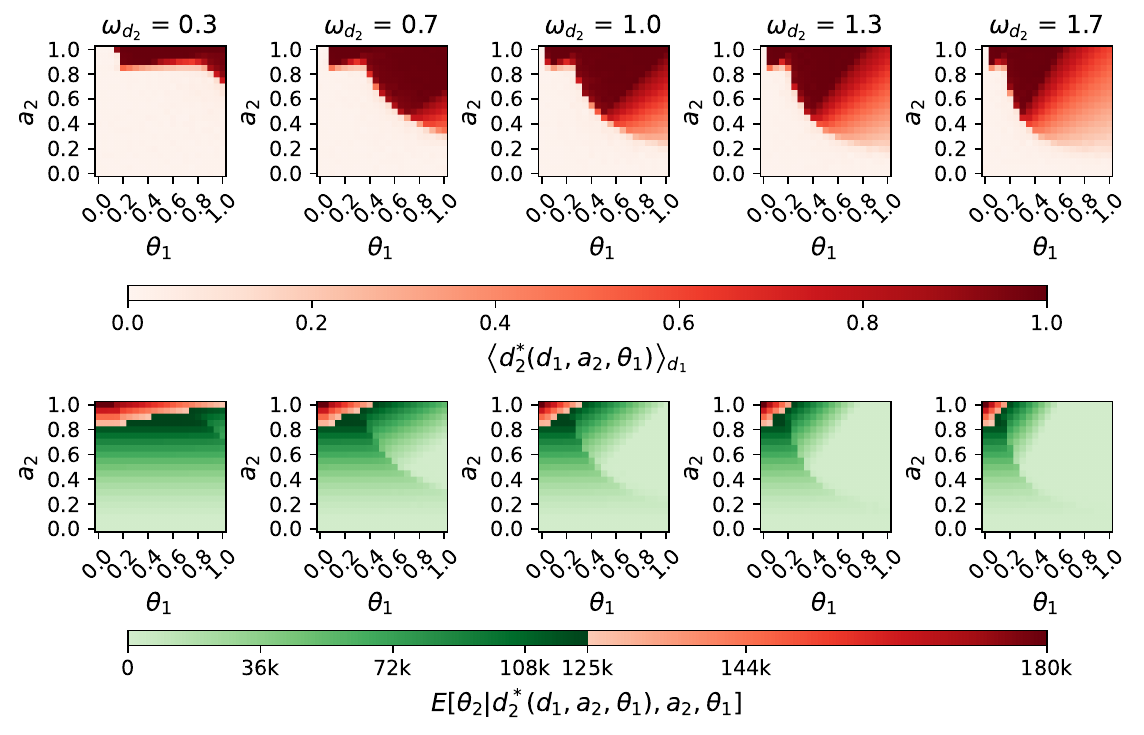}
    \vspace{-0.3cm}
    \caption{\small (Top row) Mean $\langle d_2^*(d_1,a_2,\theta_1) \rangle_{d_1}$ of \( d_2^*(d_1,a_2,\theta_1) \) over \( d_1 \) for different $\omega_{
    d_2}$. (Bottom row) Expected number 
    \( E[\theta_2 \mid d^*_2(d_1,a_2,\theta_1),a_2,\theta_1] \)
        of individuals to be infected  for different $\omega_{
    d_2}$. Best viewed in color.} 
    \label{fig:colormap_D2_theta2_sens_wd2}
\end{figure}

In the original setting, $\omega_{d_2} = 0.9$, suggesting that $a_2$ is more effective than $d_2$ and $\theta_1$. We consider scenarios where $d_2$ and $\theta_1$ have the same effect than $a_2$ ($\omega_{d_2} 
 =1$), as well as scenarios where $d_2$ and $\theta_1$ are more ($\omega_{d_2} \in \{1.3, 1.7\}$) or less ($\omega_{d_2} \in \{0.4, 0.7\}$) influential.  Figure \ref{fig:colormap_D2_theta2_sens_wd2} displays the mean of $d_2^*(d_1, a_2, \theta_1)$ over the values of $d_1$ and $E[\theta_2 \mid d^*_2(d_1, a_2, \theta_1), a_2, \theta_1]$ across such $\omega_{d_2}$ values.

 \noindent Observe the differences across different values of $\omega_{d_2}$. In the top row, as $\omega_{d_2}$ increases, there are more $(\theta_1, a_2)$ values where $d_2^{*}(d_1,a_2,\theta_1) \neq 0$ (non-white regions), meaning that $D$ should implement reactive measures to counteract $a_2$ in more instances. As $d_2$ becomes more effective, $D$ is more inclined to invest, 
  being more likely to successfully stop the disinformation spread.
   Conversely, when $\omega_{d_2}$ is lower, $ d_2^{*}(d_1,a_2,\theta_1) \approx 1$ (dark red) is applied for higher values of $\theta_1$, as maximum reactive measures are used primarily in cases where the consequences of $a_2$ may be severe, despite $d_2$ being less effective. Additionally, the regions where $0 < d_2^{*}(d_1,a_2,\theta_1) < 1$ (pale to medium red region) expand as $\omega_{d_2}$ increases, since greater effectiveness allows $D$ to better calibrate its $d_2$ investment,
    avoiding both over- and underspending. In the bottom row,  observe that, as $\omega_{d_2}$ increases, the number of cases requiring patient transfer to the private system (red region) decreases. Additionally, the expected number of infected people
     for each $(\theta_1, a_2)$ also decreases, given that light green regions expand. This occurs because, as $\omega_{d_2}$ increases, $d_2$ becomes more effective, leading to better outcomes for a same investment level.

\subsubsection{Analysis of effect of $t_d/t_a$ on the Attacker's second stage decision}

   This ratio determines the effectiveness of $d_1$ compared to 
    $a_1$ on detecting $a_2$. For fixed $d_1$, $a_1$, and $a_2$, if $t_d > t_a$, the 
    detection degree $\theta_1$ on average will be higher than if $t_d = t_a$ or $t_d < t_a$, since the investment in $d_1$ is more efficient than the investment in $a_1$. Similarly, if $t_d = t_a$, the expected value of $\theta_1$ ($E[\theta_1|d_1,a_1,a_2]$) will be higher than when $t_a > t_d$.

\begin{figure}[h]
    \centering    \includegraphics[width=0.9\textwidth]{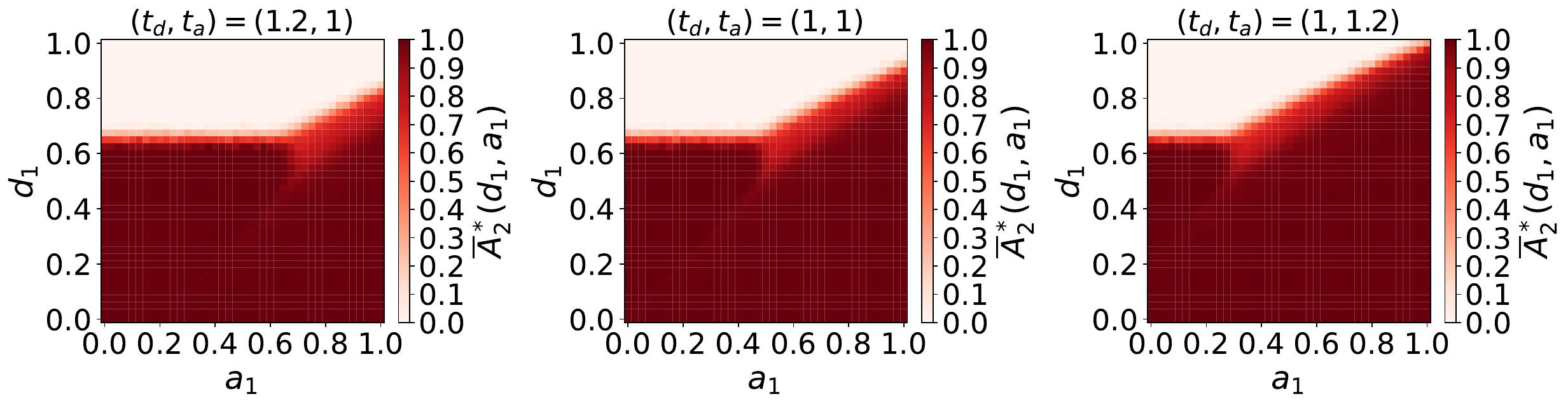}
    \caption{\small $\overline{A}_2^*(d_1,a_1)$ for several ($t_d, t_a$). Best viewed in color.}
    \label{fig:colormap_A2_sens_td_ta}
\end{figure}

 In the original setting, $t_a = 1.2$ and $t_d = 1$, suggesting that $A$'s investment is more effective than $D$'s. Let us test scenarios where the investments are equally effective ($t_d = t_a = 1$) and $D$'s investment is more effective ($t_d = 1.2$, $t_a = 1$). Figure \ref{fig:colormap_A2_sens_td_ta} displays estimates of 
$\overline{A}_2^*(d_1,a_1)$ for the different $(t_d,t_a)$ pairs.

\noindent Although the three plots display similar patterns, we observe some differences between them.  First, when $t_a/t_d$ decreases, the region 
  $\overline{A}_2^*(d_1,a_1) \approx 0$ (white) expands, suggesting that $A$ perceives a lower probability of successful outcome, thereby choosing not to attack. 
   Besides, the region where $0<\overline{A}_2^*(d_1,a_1)<1$ (moderate red) shrinks as $t_a/t_d$ decreases: this occurs because, as $A$ recognizes that the attack is less effective, the only way to compensate it is by launching a maximum attack.  Additionally, $\overline{A}_2^*(d_1,a_1) \approx 1$ (dark red) for more points $(d_1,a_1)$ as the ratio $t_a/t_d$ increases:
     $A$ is more likely to implement a higher-intensity attack when it becomes more effective, as his expected outcome is greater.   

\section{Discussion}

We have introduced a methodology that employs APS to address from an ARA perspective general security games represented as BAIDs with incomplete information.  It provides a unified approach capable of resolving any game represented as a proper BAID, irrespective of the number of decisions per agent across both discrete and continuous domains, thus being the first general solution to the query posed by \cite{Shachter1986}.  The methodology is exemplified through a disinformation war example.

The proposed approach will be particularly relevant %
in the emergent field of Adversarial Machine Learning (AML) \citep{rios2023adversarial} as the creation of scalable algorithmic approaches is imperative to address the inherent challenges in this domain. Furthermore, the deployment of ARA solutions in such contexts is essential as the usual common knowledge assumptions typically used in AML are frequently invalid, thus requiring a framework that more accurately handles uncertainty. 

However, while the methodology resolves general games across continuous domains, it may become computationally burdensome when applied  in scenarios with a large number of decision stages per agent, as it will require approximating the (random) optimal value function of the Defender (Attacker) multiple times. This computational load could be partly reduced, as several stages of the algorithm are amenable to parallelization. In particular, for each value of the grid used in Sections \ref{sec:storeD} (\ref{sec:appr_A_PSI_attack}), the Defender's optimal values (Attacker's distribution) can be approximated in parallel. Similarly, computing the MC approximation of $\hat{\psi}_D$ ($\hat{\Psi}_A$) for each value in the grid in each Defender's (Attacker's) stage can be performed in parallel. Additionally, there may be a trade-off between the explainability and performance of the models used to approximate agents' value functions. Some models, such as neural networks, may better estimate the value functions compared to less complex models like linear regression; however, a linear regressor allows for a better understanding of variable importance  in decision support as it is interpretable, in contrast to a neural network. Finally, notice that we approximate functions and conditional distributions using neural networks trained on samples obtained over fixed grids in the input space. While this ensures broad coverage, it can be computationally inefficient. A possible direction for future work is the use of adaptive or optimal experimental design to select input points more strategically. By focusing sampling on regions of high variation or model uncertainty, such methods could improve approximation quality while reducing the number of required evaluations.

Additionally, future work includes extending the presented methodology from two to several agents, addressing the computational 
  and conceptual challenges that may arise from the added complexity of handling more agents. Besides, exploring the use of Hamiltonian Monte Carlo methods for APS represents a promising avenue for future research.

\vspace{1cm}

\noindent \textbf{\textit{Credit author statement}}.
J.M.\ Camacho: Software, Writing-Original Draft, Writing-Reviewing and Editing. R.\ Naveiro: Methodology,  Writing-Original Draft, Writing- Reviewing and Editing. D. R. Insua: Conceptualization, Methodology,  
Writing-Original Draft, Writing-Reviewing and Editing, Funding Acquisition.

\noindent {\textit{\textbf{Funding.}} This work was supported by the \href{http://dx.doi.org/10.13039/501100001961}{AXA-ICMAT} Chair in Adversarial Risk Analysis, the \href{http://dx.doi.org/10.13039/100015464}{EOARD}-\href{http://dx.doi.org/10.13039/100000181}{AFOSR} project RC2APD, GRANT  13324227; the European Union’s Horizon 2020 Research
and Innovation Programme under Grant Agreement No. 101021797 (STARLIGHT); Ministerio de Ciencia y Tecnología (PID2021-124662OB-I00).
JMC is supported by a fellowship from ``la Caixa" Foundation (ID 100010434) whose code is LCF/BQ/DI21/11860063. The funding sources have not been involved in the study design, the analysis of the experiments, the writing of the work, or the decision to submit the article for publication.

\noindent {\textbf{\textit{Declarations of interest:}}  None.

\appendix

\section{Proofs of key propositions}
\label{sec:app_proof_pro}

%

\paragraph{Proof of Proposition 1} 
For each $d$ and $\omega\in \Omega$, given 
the assumptions about $U_A(a, \theta)$,  
$P_A(\theta \vert d,a) $ and ${\cal A}$, we have:
1) $\Psi ^\omega _A (d, a) =\int U^{\omega }_A (a, \theta ) P_A ^{\omega}(\theta | d, a)$ is (a.s)  continuous in $a$;
 therefore, 2)  \textcolor{white}{'}there exists a.s.\
 $A^{*} = \argmax_{x\in \mathcal{A}} \int
\Psi ^\omega _A (d, a)P_{A}^{w}(d) \dd d$ as we are maximising an a.s.\textcolor{white}{'}continuous function in a compact domain;
3) the distribution 
$ \Pi_A^{\omega} (a, \theta, d)$ 
$\propto U_A^{\omega }(a, \theta)P_A^{\omega }(\theta \vert d,a)P_A^{\omega }(d)$
is well-defined being $U^{w}_A(a,\theta)$ a.s.\textcolor{white}{'}positive. By construction, through sampling $u_A(a, \theta) \sim U_A(a, \theta)$, $p_A(\theta \vert d, a) \sim P_A(\theta \vert d, a)$ and $p_A(d)\sim P_A(d)$, one builds $\pi_A(a,\theta ,d) \propto u_A(a, \theta) p_A(\theta \vert d, a)p_A(d)$, which is a sample from $\Pi_A (a, \theta, d)$. Using \cite{smith1993bayesian} convergence results for MH algorithms, the samples generated in Algorithm \ref{alg:ARA_APS_framework}c) define a Markov chain with  $ \pi_A(a, \theta, d)$ as stationary distribution. Once convergence is detected, a consistent estimator \citep{Parzen} of the modes of the $a$ samples 
  can be built
which converges a.s.\ to $\text{mode}(\pi_A(a))$, providing a sample from $A^{*}$ whose distribution is $\Pbb_F \left[ A^*  \leq a  \right] = p_D(A\leq a)$. Then, the samples generated through the  \texttt{AAPS} function are distributed a.s$.$ according to $p_D(a)$.

As ${\cal D}$ is compact 
and $\psi_D (d)$ is continuous in $d$,  $d^\opt_\text{ARA}$ exists.
Since  $u_D$ is positive and integrable, 
$\pi_D(d,a,\theta) \propto u_D(d, \theta)p_D(\theta \vert d,a)p_D(a)$ is  
well-defined,  
and is the stationary distribution of the MH Markov chain of the $\texttt{DAPS}$ routine in Algorithm \ref{alg:ARA_APS_framework}b), using arguments in \cite{smith1993bayesian}.
Once convergence is detected,
 the marginal samples in $d$ are approximately distributed as $\pi_D(d)$,
 from which we approximate a.s.
$d^\opt_\text{ARA}$ through a consistent mode 
estimator \citep{Parzen}. $\hfill \square$

\paragraph{Proof of Proposition 3}  Let $D$ be the last Defender's decision node and  $\psi_D(y)$ be the current utility where $y$ represents the instantiations of the current antecessors of the value node. By assumption, 
  the initial $\psi_D(y)$ is positive (we prove below
  that, in the other stages, it is positive by construction). Let $\mathcal{X}_c(D) = \{X_1, \ldots, X_{n_D}\}$ be the chance nodes to be eliminated before $D$, detectable applying \cite{Shachter1986}'s algorithm qualitatively, possibly including several arc inversions, and $ant(X_i)$ designate the antecessors of node $X_i$. Then, by iteratively applying Shachter's operations, we eliminate the nodes in $(D, \mathcal{X}_c(D))$ by maximising expected utility, with the value node inheriting the antecessors of $\mathcal{X}_c(D)$ not eliminated in the process, that is  $y' = \Big(y \setminus \big(\mathcal{X}_c(D)\cup D\big)\Big)\cup ant\big(\mathcal{X}_c(D)\big)$ (and the optimal conditional decision to be stored 
$d^{*}(y')$). 

Consider the AD 
\begin{equation*}
\pi(d, x_1, \ldots, x_{n_D} | ant(\mathcal{X}_c(D))) \propto \psi_D(y) \prod_{X_i\in \mathcal{X}_c(D)} p_D(x_i|ant(X_i)),
\end{equation*}
\noindent which is well-defined as $\psi_D(y)$
is positive. If all nodes participating in arc inversions are eliminated, 
  then $d^{*}(y') =  \text{mode}(\pi(d|ant(X_c(D))))$, since the conditional marginal is proportional to the expected utility. Otherwise, the AD is proportional to the expected utility, except for a positive constant, and, again, $d^{*} (y') = \text{mode}(\pi(d|ant(X_c(D))))$. The new evaluation is 

\begin{equation*}
\psi_D(y') = \int\cdots\int \psi_D(y) \prod_{X_i\in \mathcal{X}_c(D)} p_D(x_i|ant(X_i)) \dd x_1 \ldots  \dd x_{n_D}, 
\end{equation*}

\noindent which is positive by construction.
$\hfill \square$

\paragraph{ Proof of Proposition 4} This proof is analogous to that of Proposition 3, and is therefore omitted and included in the 
 SM. $\hfill \square$

\paragraph{Proof of Proposition 5}  

 If the BAID is proper, we can define the corresponding ADP and DDP. Each time we find the corresponding last $D_i$ in the DDP.
\begin{itemize}
    \item 
  If all $p_D$'s are available for reduction, we reduce node $D_i$ and eliminate it from the DDP
    using Proposition 3, with the utility model being 
    proportional to the expected utility model, therefore leading to 
    the same optimal decision.
 \item    
 If not, we find the required $p_D$'s through \texttt{AAPS\_red} operations, eliminating the corresponding nodes from the ADP, using Proposition 4,
    possibly several times. Then, reduce the corresponding $D_i$ from the DDP, with the same 
    optimality preserving features. 
\end{itemize}

 \noindent As both the DDP and the ADP are finite, the algorithm necessarily terminates providing the required output as deduced from Propositions \ref{prop:DAPS_red} and \ref{prop:AAPS_red}. $\hfill \square$

\section{List of abbreviation.}

To facilitate reading, we provide a summary of the abbreviations used in the manuscript.

\begin{table}[h!]
\begin{tabular}{cccc}
\hline
DHS  & Defense and Homeland Security                                                     & ARA & Adversarial Risk Analysis        \\ \hline
ID   & Influence Diagram                                                                 & APS & Augmented Probability Simulation \\ \hline
BAID & Bi-Agent Influence Diagrams                                                       & AD  & Augmented Distribution           \\ \hline
MCMC & Markov Chain Monte Carlo                                                          & MH  & Metropolis-Hastings              \\ \hline
RAD  & Random Augmented Distribution                                                     & DC  & Disinformation campaign          \\ \hline
RAPS & \begin{tabular}[c]{@{}c@{}}Random Augmented Probability\\ Simulation\end{tabular} & DDP & Defender Decision Path           \\ \hline
ADP  & Attacker Decision Path                                                            & NN  & Neural Network                   \\ \hline
SM   & Supplementary Materials                                                           & MLP & Multilayer Perceptron            \\ \hline
NLL  & Negative log-likelihood                                                           & MC  & Monte Carlo                      \\ \hline
AML  & Adversarial Machine Learning                                                      &     &                                  \\ \hline
\end{tabular}
\caption*{\small Table 2: Abbreviations in the manuscript.} 
\end{table}

\bibliography{paper-refs} %

\pagebreak

\section*{{\large Supplementary materials: Computational adversarial risk analysis for general security games.}}

These supplementary materials include the proof of Proposition 4 from the main text (Section SM1), the parameters describing the disinformation war case study (Section SM2), the chosen augmentation parameters $h$ to increase APS's efficiency (Section SM3), details to select the models estimating $\psi_A(d_1, a_2, \theta_1)$ (Section SM4), $p_D(a_2|d_1, a_1)$ (Section SM5), and $\Psi_A(d_1, a_1)$ (Section SM6), as well as the specification of the distribution approximating $p_D(a_1)$ (Section SM7).

\section*{SM1. Proof of Proposition 4.}

Let $A$ be the last Attacker's decision node and  $\Psi_A(y)$ be the current random utility where $y$ represents the instantiations of the current antecessors of the value node, which includes $D$. By assumption, the initial $\Psi_A(y)$ is a.s.\textcolor{white}{'}positive (we prove below by construction that it will also be a.s.\textcolor{white}{'}positive for earlier nodes). Let $\mathcal{X}_c(A) = \{X_1, \ldots, X_{n_A}\}$ be the chance nodes to be eliminated before $A$ using \cite{Shachter1986}'s algorithm, possibly including several arc inversions, and $ant(X_i)$ the antecessors of node $X_i$. By applying iteratively \cite{Shachter1986}'s operations, we eliminate the nodes in $(A, \mathcal{X}_c(A))$ by maximizing the random expected utility with the value node inheriting the antecessors of $\mathcal{X}_c(A)$ not eliminated in the process, that is $y' = \Big(y \setminus \big(\mathcal{X}_c(A)\cup A\big)\Big)\cup ant\big(\mathcal{X}_c(A)\big)$ and the random optimal decisions to be stored are $A^{*}(y')$.  

Consider the RAD

\begin{equation*}
\Pi(a, x_1, \ldots, x_{n_A} | ant(\mathcal{X}_c(A))) \propto \Psi_A(y) \prod_{X_i\in \mathcal{X}_c(A)} P_A(x_i|ant(X_i)),
\end{equation*}
\noindent which is a.s.\textcolor{white}{'}well-defined as 
 \textcolor{white}{'}$\Psi_A(y)$
is positive a.s.
If all nodes participating in arc inversions are eliminated, then $A^{*}(y') = \text{mode}(\Pi(a|ant(X_c(A))))$, since the random conditional marginal is a.s.\textcolor{white}{'}proportional to the random expected utility. Otherwise, the RAD is a.s.\textcolor{white}{'}proportional to the random expected utility, except for a positive constant and, therefore, again $A^{*}(y') = \text{mode}(\Pi(a|ant(X_c(A))))$. The new random evaluation is defined as

\begin{equation*}
\Psi_A(y') = \int ... \int  \Psi_A(y) \prod_{X_i\in \mathcal{X}_c(A)} P_A(x_i|ant(X_i)) \dd x_1 \ldots  \dd x_{n_A},
\end{equation*}
which is a.s.\textcolor{white}{'}positive.
$\hfill \square$

\pagebreak

\section*{SM2. Modeling details in case study} \label{sec:app_numerical_ex}

This section describes the modeling assumptions made by the Defender to numerically solve the disinformation war case study in Section \ref{sec:casestudy}.


\subsection*{SM2.1 D's problem}

Recall $D$'s problem, depicted in the ID in Figure \ref{fig:app_D_problem_ID}.

\begin{figure}[h!]
    \centering    \includegraphics[width=0.32\textwidth]{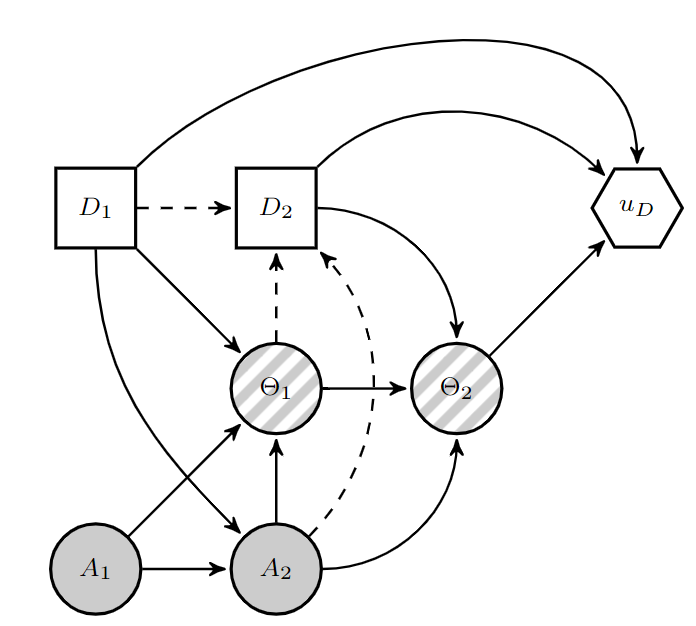}
    \caption{\small ID representing D's problem. }
    \label{fig:app_D_problem_ID}
\end{figure}

\noindent Initially, $D$ must determine the proportion of resources $d_1 \in [0,1]$ available to invest in proactive protection against disinformation. Then, after observing $\theta_1 \in [0,1]$, $D$ allocates reactive resources $d_2$. The maximum budgets that $D$ can allocate to $D_1$ and $D_2$ are $b_{d_1} = \$400$M and $b_{d_2} = \$200$M, respectively. $A$ allocates resources $a_1 \in [0,1]$ from its budget to improve its capabilities prior to the attack, assuming a maximum budget of $b_{a_1} =~\$ 380M$. Then, in the second stage, $a_2 \in [0,1]$ represents the intensity of the attack; the higher the value of $a_2$, the greater the intensity.

The interaction between $D$ and $A$'s decisions gives rise to random events $\Theta_1$ and $\Theta_2$. The first one is interpreted as the degree of recognition of the attack by $D$, influencing the effectiveness of $d_2$. We assume that
\begin{equation}
\Theta_1|d_1,a_1,a_2 \sim Be (\tau_1, \tau_2); \quad \tau_1 = \alpha_{\theta_1}\mu_{\theta_1}(\phi_{\theta_1} + \epsilon),  \quad \tau_2 =\alpha_{\theta_1}(1-\mu_{\theta_1})(\phi_{\theta_1} + \epsilon),
\label{eq:s1_dist}
\end{equation}
\noindent where $\mu_{\theta_1}$ denotes the mean of $\Theta_1$, $\phi_{\theta_1} = \max\left(\frac{1}{\mu_{\theta_1}},\frac{1}{1-\mu_{\theta_1}}\right)$, $\alpha_{\theta_1}\geq 1$ is a coefficient that helps to control the variance and $\epsilon>0$ forces the probability density function to be concave. In our problem, we consider $\alpha_{\theta_1}=2$ and $\mu_{\theta_1} = \min\Big(a_2 \times \frac{d_1 + t_d}{a_1 + t_a},1-\delta\Big)$ with \( t_d, t_a \geq 1 \) and sufficiently small $\delta>0$. The ratio \( t_d/t_a \) represents the relative effectiveness of $A$’s investment compared to $D$’s. It ensures that, even when \( d_1 = 0 \) and \( a_1 = 0 \), the recognition degree (\( \theta_1 \)) may still be high due to \( a_2 \). This follows from our assumption that for large values of \( a_2 \), \( \theta_1 \) will remain significant. As \( a_2 \) reaches a substantial portion of the population, $D$’s recognition will also increase.  The extent of this effect depends on \( t_d \) and \( t_a \). In our case, we set \( t_d = 1 \) and \( t_a = 1.2 \). Since \( t_a > t_d \), $A$’s investment is more effective than $D$’s when \( d_1 \approx a_1 \), meaning that $D$ must invest more to counteract $A$’s efforts. Observe that \( \mu_{\theta_1} = 0.83 \) when \( d_1 = 0 \), \( a_1 = 0 \), and \( a_2 = 1 \). Furthermore, if $a_2 = 0$, we assume that $\theta_1 = 0$, as no attack has been executed. On the other hand, $\delta$ is used to avoid $\mu_{\theta_1} = 1$, as this value is not within the support of the Beta distribution.

$\Theta_2$ designates the number of individuals affected by the disinformation campaign, modelled as
\begin{equation}
\Theta_2 |d_2,a_2,\theta_1 \sim Bin(\phi_1, \phi_2); \quad \phi_1 = \lfloor a_2 \times n\rfloor, \quad \phi_2 = \max(0, a_2 - w_{d_{2}}\times \theta_1 \times {d_2})
\label{eq:S_2_dist}
\end{equation}
\noindent where $\lfloor \emph{x} \rfloor$ denotes the floor function applied to  \emph{x}, $n$ are the number of $D$'s citizens and 
$\omega_{d_2}>0$ regulates how effective $d_2$ and  $\theta_1$  are regarding $A$'s attack. When $\omega_{d_2}>1$, reactive measures are more effective: the attack may be stopped and have no consequences even when there is no full recognition ($\theta_1 \neq 1$) and maximum resources are not allocated ($d_2\neq 1$). Instead, when $\omega_{d_2}<1$, it means that it may be possible that using all resources for reactive measures ($d_2=1$) and full recognition ($\theta_1 =1 $), it would be impossible to stop the attack spread. In the example, $n = 180000$ and  $\omega_{d_2}=0.9$, implying that when $a_2 = 1$, the defenses would be only able to reduce the probability $\phi$ up to 0.1.

Finally, consider the utility function $u_D(d_1, d_2, \theta_2)$.  Define first a value function $v_D$ aggregating monetarily the outcomes

\begin{itemize}
    \item  The minimal security program against disinformation $(d_1 = 0)$ entails a cost of $d_1^0=15M$. Therefore, the cost of $D_1$ will be  $m_1^D(d_1) =d_1 \times b_{d_1}+  d_1^0$.

\item  The cost of $D_2$ corresponds directly to the investment, that is $m_2^D(d_2) = d_2 \times b_{d_2}$.

 \item Let $r$ denote the cost of having an individual in the hospital until its recovery, and $c$ the maximum capacity of $D$'s health system. Once capacity is reached, the health system is unable to accommodate more patients, and infected citizens will need to be transferred to private hospitals for treatment, incurring in additional costs. Thus,
\begin{equation*}
   m_s^{D}(\theta_2)=
    \begin{cases}
        r \times \theta_2   & \text{if } \theta_2 \leq c \\
        r \times \theta_2 + (\theta_2-c) \times l & \text{if } \theta_2 > c  ,
    \end{cases}
\end{equation*}
where $l$ corresponds to the additional costs incurred when a citizen is treated in the private health system. In 
 the example, $r = \$0.005$ M, $c= 125000$ and $l = \$ 0.02$ M.

\end{itemize}

\noindent Then, the value function $v_D$ is

\begin{equation*}
v_D(d_1,d_2, \theta_2) = -m_1^{D}(d_1) - m_2^{D}(d_2) - m_s^{D}(\theta_2).
\end{equation*}

\noindent  For simplicity, assume that, being risk neutral, 
$D$'s utility is linear in costs,

\begin{equation*}
u_D(d_1,d_2,\theta_2) = v_D(d_1, d_2, \theta_2) + \gamma_D,
\end{equation*}

\noindent where $\gamma_D$ is a constant such that $ u_D > 0$, to ensure APS requirements. We used $\gamma_D(=2616)$.

\pagebreak

\subsection*{SM2.2 A's problem}

Recall the problem faced by $A$, displayed in the ID in Figure \ref{fig:app_A_problem_ID}.

\begin{figure}[h!]
    \centering    \includegraphics[width=0.32\textwidth]{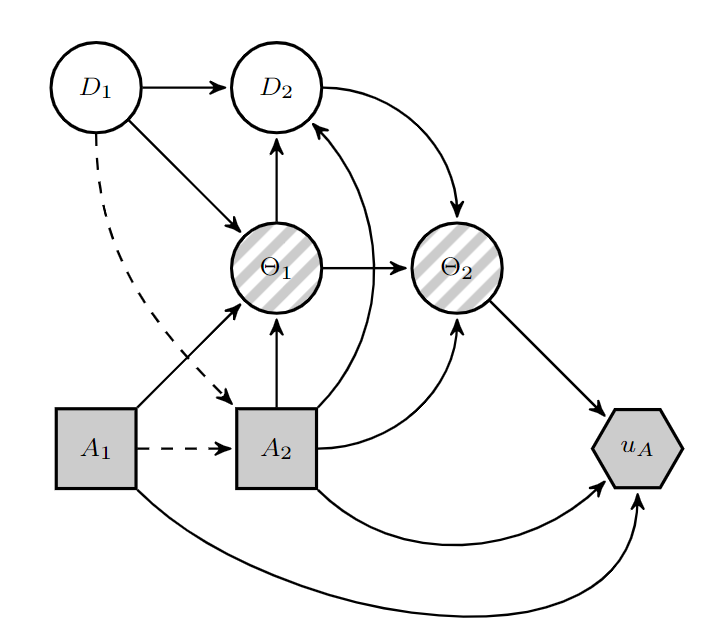}
    \caption{\small ID representing A's problem. }
    \label{fig:app_A_problem_ID}
\end{figure}

\noindent To evaluate $A$'s estimates on random events, $D$ assumes that $A$'s assessment will correspond to hers with some additional uncertainty. As relevant data about the scenario becomes available, $D$ could update her estimates in a Bayesian manner.

Considering that $\Theta_1$ is distributed as in  (\ref{eq:s1_dist}), we incorporate uncertainty regarding $\tau_1$ and $\tau_2$ through

\begin{equation*}
\Theta_1|D_1, A_1, A_2 \sim Be (\mathcal{T}_1, \mathcal{T}_2); \quad \mathcal{T}_1 = \kappa \times \tau_1 , \mathcal{T}_2 = \kappa \times \tau_2
\end{equation*}

\noindent where $\kappa \sim \mathcal{U} \left(\frac{3}{5},\frac{3}{4}\right)$. As a result, the variance of $P_A(\Theta_1| D_1, A_1, A_2)$ is bigger than that of $p_D(\theta_1| d_1, a_1, a_2)$, while $\mu_{\theta_1}$ remains the same.

Based on (\ref{eq:S_2_dist}), $\Theta_2$ 
is modelled as
\begin{equation*}
\Theta_2|\Theta_1, D_2, A_2\sim Bin\big(\phi_1,\Phi_2),
\end{equation*}
\noindent where we do not consider uncertainty regarding $\phi_1$, as it is reasonable to assume that both $D$ and $A$ have a similar estimate of the number of individuals susceptible 
of being affected by the disinformation. Otherwise, $\Phi_2\sim \max\Big(0, \min\big(1,\mathcal{U}\big((1-\delta_{\phi_2})\times\phi_2, (1+\delta_{\phi_2})\times\phi_2)\big)$\big)\Big). Thus, $\Phi_2$ is selected uniformly at random within an interval of width $2\times\delta_{\phi_2}\times\phi_2$ centered around $\phi_2$: $D$ assumes that $A$'s estimates may differ by up to $\delta_{\phi_2}\% $ from its own, with higher $\delta_{\phi_2}$ indicating greater uncertainty. We set $\delta_{\phi_2}=0.05$, allowing for up to $5\%$ difference from the real value.

We next assess $P_A(D_1=d_1)$ and $P_A(D_2|D_1,A_1,\Theta_1)$. To prevent recursive loops, we regard both probabilities as based on expert knowledge, allowing for a certain level of uncertainty. We consider that

\begin{itemize}
    \item For $D_1$, the defender assumes that $A$ estimates she will spend approximately 70$\%$ of $b_{d_1}$ (budget for $D_1$) on proactive defense. We model $P_A(D_1)\sim Be\big(\kappa \times \mu_{d_1}, \kappa \times (1-\mu_{d_1})\big)$ where $\mu_{d_1}=0.7$ and $\kappa\sim\mathcal{U}(7.5,8)$.
    \item $D$ supposes that $A$ models her second decision as 
\begin{equation*}
p_A(d_2|d_1,\theta_1,a_2)\sim Be(v_1,v_2); \quad \upsilon_1 = \alpha_{d_2}\mu_{d_2}(\phi_{d_2} + \epsilon), \quad \upsilon_2 = \alpha_{d_2}(1 - \mu_{d_2})(\phi_{d_2} + \epsilon)
\end{equation*}
\noindent where $\phi_{d_2} = \max(\frac{1}{\mu_{d_2}},\frac{1}{1-\mu_{d_2}})$, $\alpha_{d_2} = 2$ is a coefficient to regulate the variance, and $\epsilon>0$, as before, to force the distribution to be concave. In our problem, $\mu_{d_2}= \min (\frac{\theta_1 a_2}{a_2 + (1-d_1)}, 1-\delta)$, with sufficiently small $\delta>0$. As before, $\delta$ is used to avoid $\mu_{d_2}=1$, as it is not within the Beta distribution support. With this parametrization, it is assumed that if there is no prior preparation (\( d_1 = 0 \)), but maximum recognition (\( \theta_1 = 1 \)) and maximum attack (\( a_2 = 1 \)), the estimated mean of \( D_2 \)  would be 0.5.  Additionally, for fixed \( a_2 \) and \( d_1 \), a higher recognition level (\( \theta_1 \)) will result in a higher mean of \( \mu_{d_2} \). Similarly, for fixed \( a_2 \) and \( \theta_1 \), greater proactive research (\( d_1 \)) leads to a higher \( u_{d_2} \).  Moreover, for fixed values of \( d_1 \) and \( \theta_1 \), an increase in \( a_2 \) also increases \( \mu_{d_2} \). Furthermore, assume that $d_2 = 0$ when (a) $a_2 = 0$, as no attack has been produced and, therefore, no defense is necessary; (b) $\theta_1 = 0$, given that $D$ does not perceive any attack and thus does not execute any countermeasure. 

Then, uncertainty is included in  $\upsilon_1$ and $\upsilon_2$ through 

\begin{equation*}
P_A(D_2|D_1,\Theta_1, A_2) \sim Be (\Upsilon_1, \Upsilon_2); \quad \Upsilon_1 = \kappa \times \upsilon_1, \quad \Upsilon_2 = \kappa \times \upsilon_2
\end{equation*}

where $\kappa \sim \mathcal{U} \left(\frac{3}{5},\frac{3}{4}\right)$. As a result, the variance of $P_A(D_2|D_1,\Theta_1,A_2)$ increases compared to $p_A(d_2|d_1,\theta_1,a_2)$, while $\mu_{d_2}$ remains the same.

\end{itemize}

Finally, $D$ assesses $A$'s random utility $U_A(a_1, a_2, \theta_2)$.  Consider a scenario where the defender, through expert judgment, has developed an understanding of the attacker's value function. In particular, assume that

\begin{itemize}

\item For $A_1$, the cost will correspond to his investment in improving its attack without accounting for uncertainty, i.e, $m^A_1(a_1) = a_1 \times b_{a_1}$.

\item Without loss of generality,  consider that the intensity of an attack on $A_2$ is linear in costs. The costs to account correspond to the expenses to be paid to disseminate the disinformation, such as paying social media platforms to propagate the content through ads.  Therefore, we model the cost of the intensity as: $M_2^A(a_2) = Y_2^A \times a_2$, where $Y_{2}^A\sim \mathcal{U}\big((1-\delta_{y_2^A})\times y_2^A ,(1+\delta_{y_2^A})\times y_2^A\big)$, with $y_2^A = \$ 300M$ and $\delta_{y_2^A} = 0.05$.

\item Suppose the assessment of $A$ on $\theta_2$ would be similar to $D$'s, having $A$ uncertainty about $D$ on the cost of treating a patient, $r_{1}^{A}$, and the capacity of $D$'s health system, $c^A$. $D$ assumes that $A$ considers treating a person costs $r^A_1 = \$0.005M$; to account for uncertainty, it will be modeled as $R_1 \sim \mathcal{U}\big((1-\delta_{r_1^{A}})\times r_1^A, (1+\delta_{r_1^{A}})\times r_1^A\big)$. On the other hand, $D$ estimates that $A$ believes that the health system has the capacity to treat up to $c^A$ = 125000 patients, and the extra costs per patient to be transferred to the private system to be $l^A= \$0.02 M$. Similarly, to account for uncertainty, the capacity of the health system and the extra costs of transferring a patient will be modeled as $C^A\sim\mathcal{U}\big((1-\delta_{c^A})\times c^A, (1+\delta_{c^A})\times c^A\big)$ and $L\sim \mathcal{U}\big((1- \delta_l^A)\times l^A,(1+\delta_l^A)\times l^A\big)$, respectively.  Therefore, $M_{s}^{A}(\theta_2)$ is modelled as 

\begin{equation*}
   M_s^{A}(\theta_2)=
    \begin{cases}
        R_1 \times \theta_2   & \text{if } \theta_2 \leq C^A \\
        R_1\times \theta_2 + (\theta_2 - C^A)\times L & \text{if } \theta_2 > C^A
    \end{cases}
\end{equation*}

\end{itemize}

\noindent In the case study, $\delta_{r_1^A} = 0.05$, $\delta_{c^A} = 0.05$ and $\delta_{l^A} = 0.01$, assuming that $A$ may have accurate estimates based from its intelligence services. Combining the previous assumptions,

\begin{equation*}
V_A(a_1,a_2,\theta_2) = - m_1^A(a_1) - M^A_2(a_2) + M^A_s(\theta_2)
\end{equation*}

\noindent Finally, for simplicity, assume that $A$'s utility is linear in costs

\begin{equation*}
U_A(a_1,a_2, \theta_2) = V_A(a_1, a_2, \theta_2)+\gamma_A,
\end{equation*}

\noindent  where $\gamma_A $ is a  constant so that $U_A > 0$, to guarantee applicability of APS.  We used $\gamma_A (=1280)$.

\section*{SM3. Augmentation parameters $h$ used in D's and A's decision stages.}

To enhance the efficiency of APS in all decision stages, for both the Defender and the Attacker, we used the approach described in Section \ref{sec:enhance_APS}, which involves an augmentation parameter $h$. Table \ref{tab:h_aug_param} displays the parameter $h$ used at each decision stage.

\begin{table}[h!]
\centering
\small
\begin{tabular}{ccccc}
\hline
\begin{tabular}[c]{@{}c@{}}Decision\\ stage\end{tabular} & \begin{tabular}[c]{@{}c@{}}$D$'s second  \\ (Section \ref{sec:def_second_stage_decision})\end{tabular} & \begin{tabular}[c]{@{}c@{}}$A$'s second \\ (Section \ref{sec:att_second_stage_decision})\end{tabular} & \begin{tabular}[c]{@{}c@{}}$A$'s first \\ (Section \ref{sec:att_1st_dec})\end{tabular} & \begin{tabular}[c]{@{}c@{}}$D$'s first\\ (Section \ref{sec:def_2_decision})\end{tabular} \\ \hline
$h$                                                      & 40                                                                      & 80                                                                     & 120                                                                   & 20                                                                   \\ \hline
\end{tabular}
\caption{\small Augmentation parameter $h$ used in the Defender's and Attacker's decision stages.}
\label{tab:h_aug_param}
\end{table}

To determine $h$ for each stage, several values were tested heuristically until convergence was achieved. The objective was to identify a value of $h$ that ensured convergence while avoiding unnecessarily high computational cost to obtain the solution.

\section*{SM4. Approximating $\psi_D(d_1, a_2, \theta_1)$ from Defender's second stage decision (Section \ref{sec:def_second_stage_decision}).} \label{sec:app_numerical_psi_d2}

Following the process described in Section \ref{sec:results}, we select the best multilayer perceptron (MLP) from among those with $n \in \{1,2,3\}$ hidden layers, $t \in \{16,32,64\}$ neurons per hidden layer, and ReLU activation functions, using $\mathcal{D}_{{\psi}_{D_2}}$. We evaluate the performance of each MLP configuration using the mean absolute error (MAE) and root mean square error (RMSE). Table \ref{ref:res_Arch_NN} presents the mean and standard error of the cross-validation (CV) results for the five best models over 10 iterations. Based on these results, we select the model with three hidden layers containing 32, 64, and 16 neurons, respectively.

\begin{table}[h!]
\centering
{\small
\begin{tabular}{cccccc}
\hline
Architecture & {\textbf{[}}\textbf{32,64,16}{\textbf{]}}   & {[}32,64,32{]}   & {[}32,64,64{]}   & {[}16,64,64{]}   & {[}64,64,32{]}   \\ \hline
MAE            & \textbf{10.2$\pm$0.7} & 11.3$\pm$0.4 & 11.7$\pm$0.8 & 11.6$\pm$0.6 & 12.8$\pm$1.4 \\
RSME           & \textbf{12.5$\pm$0.8} & 13.2$\pm$0.6 & 13.3$\pm$0.8 & 14.1$\pm$0.7 & 14.5$\pm$1.4 \\ \hline
\end{tabular}}
\caption{\small Mean and standard error of 5-fold CV results over 10 iterations for best five MLP architectures. Each element of the architecture represents a hidden layer, with the number indicating the neurons.}
\label{ref:res_Arch_NN}
\end{table}

After selecting the architecture, we train an MLP model on the full training dataset and evaluate it on the test dataset, obtaining an MAE of 9.3 and an RMSE of 12.9. Given the scale of $\psi_D(d_1, a_2, \theta_1)$, with a mean of 2143.6 and a median of 2206.0 in $\{\psi_D(d_1, a_2, \theta_1)\}_{j=1}^J$, these metrics indicate that the proposed MLP effectively approximates the target expected utility. Finally, we train the MLP with the selected hyperparameter configuration over the entire dataset $\mathcal{D}_{\psi_{D_2}}$ to obtain $\hat{\psi}_D(d_1,a_2,\theta_1)$, used for the attacker's first-stage decision.

\section*{SM5. Approximating $p_{D}(a_2 | d_1, a_1)$ from Attacker's second stage decision. (Section \ref{sec:att_second_stage_decision})}

We approximate $A_2^*(d_1,a_1)$ for each $(d_1,a_1)$ with a mixture of Beta distributions (MoB), with at most two components $(N\in\{1,2\})$ described by the parameters $\{(w_i, \alpha_i, \beta_i)\}_{i=1}^N$. As  Section~\ref{sec:att_second_stage_decision} describes, we use an MLP that outputs $\{(w_i, \alpha_i, \beta_i)\}_{i=1}^N$ for each $(d_1,a_1)$, with a softmax activation applied to the neurons producing $\{w_i\}_{i=2}^N$ and a softplus activation applied to those producing $\{\alpha_i, \beta_i\}_{i=1}^N$. The MLP configuration is selected following the procedure outlined in Section~\ref{sec:results}, from among models with $N\in \{1, 2\}$ mixture components, $n\in \{1, 2, 3\}$ hidden layers, $t\in \{16, 32, 64\}$ neurons per layer, and ReLU activations in the hidden layers. Table~\ref{tab:results_MOB_CV} reports the mean and standard error of the CV results for the five best-performing MLP configurations across 10 iterations. Based on these results, we select the model with three hidden layers of 64 neurons each, whose outputs define a MoB with $N = 2$ components.

\begin{table}[h!]
\centering
\small
\begin{tabular}{cccccc}
\hline
Arch.;\# Comp. & {\textbf{[}}\textbf{64,64,64}{\textbf{]}}\textbf{;2} & {[}32,64,64{]};2 & {[}64,64,32{]};2 & {[}64,32,64{]};2 & {[}32,64,32{]};2 \\ \hline
NLL            & \textbf{-31.8$\pm$0.1}    & -31.7$\pm$0.1    & -31.7$\pm$0.2    & -31.5$\pm$0.2    & -31.4$\pm$0.2    \\ \hline
\end{tabular}
\caption{\small Mean and standard error of the 5-fold CV results over 10 iterations for best five MLP architectures (Arch.) and number of MoB components (\# Comp.) used to estimate $A_2^*(d_1, a_1)$.}
\label{tab:results_MOB_CV}
\end{table}

Once the configuration is selected, the MLP is trained on the complete training dataset and evaluated on the test set, yielding an NLL of -39.9. Figure \ref{fig:MOB_results_grid} (blue) displays $A_2^*(d_1,a_1)$ for some $(d_1,a_1)$ in the test set, while Figure \ref{fig:MOB_results_grid} (red) shows the MoB approximation for the same $(d_1,a_1)$.   The final MLP, trained on the full dataset $\mathcal{D}_{A_2}$, is used in the Defender’s first-stage decision process.

\begin{figure}[h!]
    \centering
    \begin{subfigure}[b]{0.3\textwidth}
        \includegraphics[width=\textwidth]{{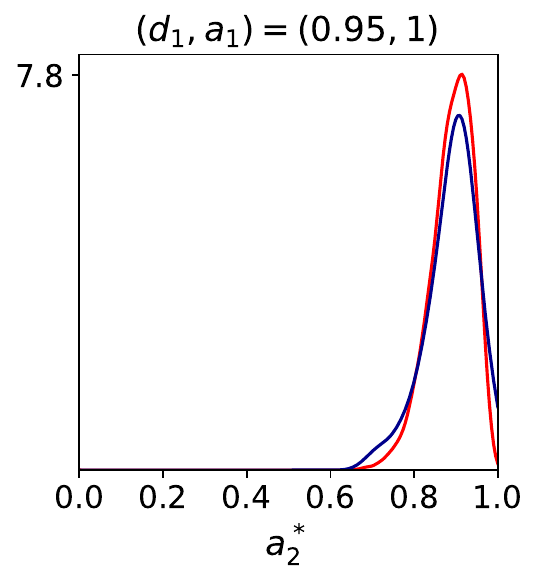}}
    \end{subfigure}
    \hfill
    \begin{subfigure}[b]{0.31\textwidth}
        \includegraphics[width=\textwidth]{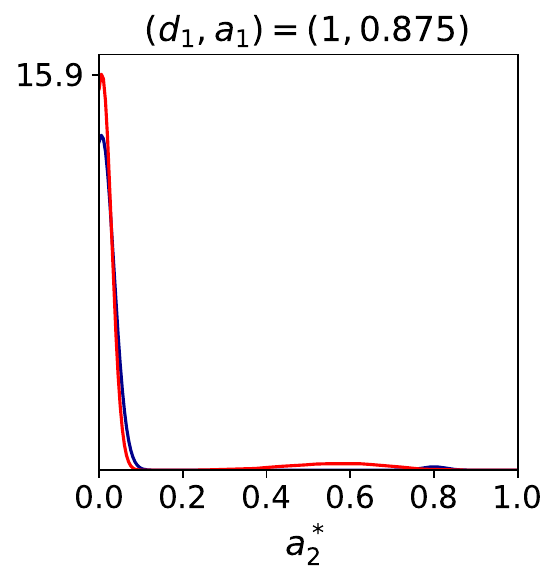}
    \end{subfigure}
    \hfill
    \begin{subfigure}[b]{0.30\textwidth}
        \includegraphics[width=\textwidth]{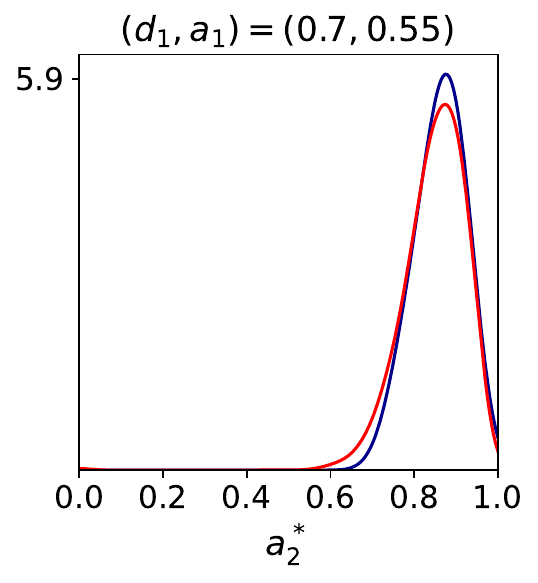}
    \end{subfigure}
    \vspace{-0.35cm}
    \caption{Empirical distribution of $A_2^*(d_1,a_1)$ (blue) against MoB approximation (red). Best viewed in color.}
    \label{fig:MOB_results_grid}
\end{figure}

\section*{SM6. Approximating $\Psi_A(d_1, a_1)$ from Attacker's second stage decision. (Section \ref{sec:att_second_stage_decision})}}

To approximate $\Psi_A(d_1, a_1)$, we use an MLP that defines a mixture of Weibull distributions (MoW) with at most $N = 2$ components. Specifically, the MLP output consists of the parameters $\{w_i, \lambda_i, \alpha_i\}_{i=1}^N$ for an input $(d_1, a_1)$, characterizing the MoW. We follow the procedure described in Section \ref{sec:results} to select the best MLP architecture from among the models with $N \in \{1, 2\}$ components, $n \in \{1, 2, 3\}$ hidden layers, and $t \in \{16, 32, 64\}$ neurons per layer. Additionally, all considered architectures use ReLU activations in the hidden layers, a softmax activation for $\{w_i\}_{i=1}^N$, and a softplus activation for $\{\lambda_i, \alpha_i\}_{i=1}^N$. Table~\ref{tab:MOW_tab_res} reports the mean and standard error of the CV results for the five best-performing MLP models over 10 iterations. In light of the results, we select the MLP with $N = 2$ and three hidden layers of 64 neurons each, obtaining a NLL of -31.7 on the test set.

\begin{table}[h!]
\small
\centering
\begin{tabular}{cccccc}
\hline
Arch.;\# Comp. & {[}\textbf{64,64,64}{\textbf{]}}\textbf{;2} & {[}64,32,64{]};2 & {[}64,64,64{]};1 & {[}64,64,32{]};2 & {[}64,64{]};2 \\ \hline
NLL            & \textbf{-25.7$\pm$0.1}    & -25.1$\pm$0.1    & -24.8$\pm$0.0    & -24.7$\pm$0.1    & -24.7$\pm$0.1    \\ \hline
\end{tabular}
\caption{ \small Mean and standard error of the 5-fold CV results over 10 iterations for the top five MLP architectures (Arch.) and number of MoW components (\# Comp.) used to estimate $\Psi_A(d_1, a_1)$.}
\label{tab:MOW_tab_res}
\end{table}

Figure \ref{fig:MoW_results} shows the MoW approximation (in red) against the probability density function density of the empirical $\Psi_A (d_1, a_1)$ (in blue) for some $(d_1,a_1)$. Finally, we train the MLP with the selected hyperparameters on the full dataset $\mathcal{D}_{\Psi_{A_2}}$ for use in the Attacker's first-stage.

\begin{figure}[h!]
    \centering
    \begin{subfigure}[b]{0.31\textwidth}
        \includegraphics[width=\textwidth]{{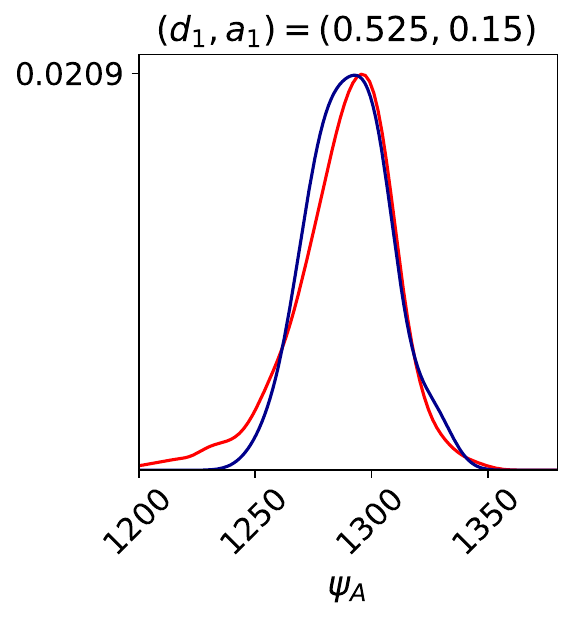}}
    \end{subfigure}
    \hfill
    \begin{subfigure}[b]{0.322\textwidth}
        \includegraphics[width=\textwidth]{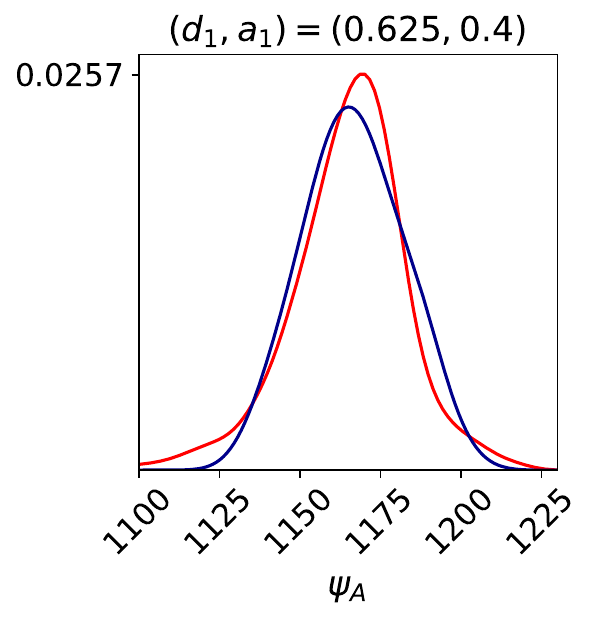}
    \end{subfigure}
    \hfill
    \begin{subfigure}[b]{0.313\textwidth}
        \includegraphics[width=\textwidth]{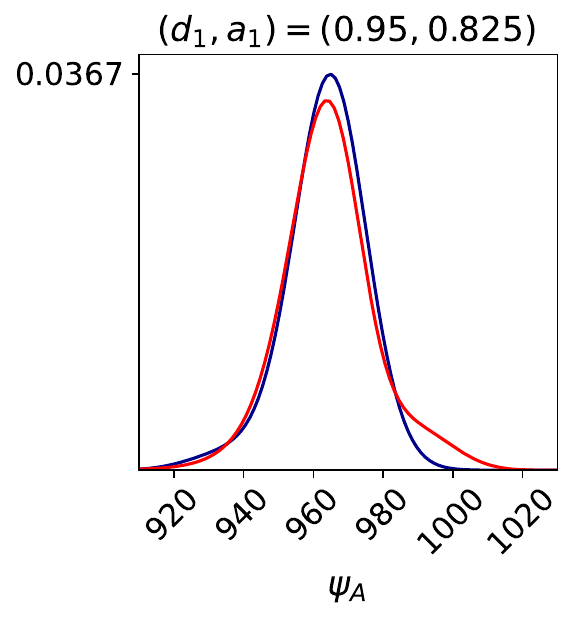}
    \end{subfigure}
        \vspace{-0.2cm}
    \caption{Density of empirical $\Psi_A(d_1,a_1)$ (blue) and MoW model approximation (red). Best viewed in color.}
    \label{fig:MoW_results}
\end{figure}

\section*{SM7. Approximating $p_D(a_1)$ from Attacker's first stage decision. (Section \ref{sec:att_1st_dec})} \label{sec:app_numerical_A2}

Given the shape of the distribution $A_1^*$ (Figure~\ref{fig:x1} blue), and the fact that it is bounded on $[0,1]$, we use 
Maximum Likelihood Estimation (Hastie et al., 2017) to fit a two-component Beta mixture model. The parameters of the mixture are $\big(\{\alpha_i\}, \{\beta_i\}, \{w_i\}\big)_{i=1}^2 = \big((100.9, 2.3), (1.6,\\ 213.4), (0.1, 0.9)\big)$.  Figure \ref{fig:x1} (red) displays 10000 samples of the fitted distribution compared to the empirical distribution of $A_1^*$ (blue), showing a good approximation.

\begin{figure}[!h]
    \centering    \includegraphics[width=0.31\textwidth]{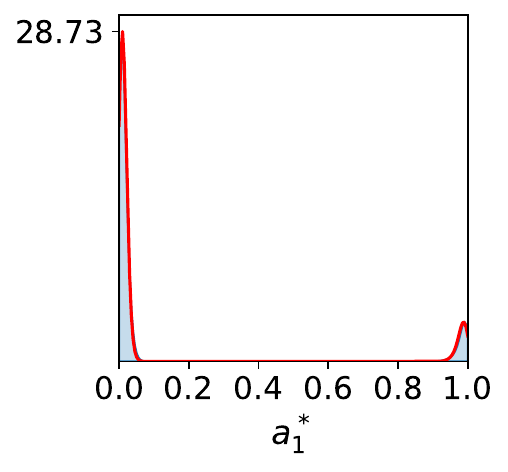}
    \vspace{-0.9em}
    \caption{\small Comparison between  empirical distribution (blue) of $A_1^*$ and its approximation (red). Best viewed in color.}
    \label{fig:x1}
\end{figure}

\pagebreak
\noindent \textbf{References}

\noindent Hastie, T., Tibshirani, R., \& Friedman, J. (2017). \textit{The elements of statistical learning: data mining, inference, and prediction.} Springer.

\end{document}